\documentclass[a4paper,11pt]{article}
\usepackage{jheppub}
\usepackage[english]{babel}

\usepackage{amsmath}
\usepackage{subcaption}
\usepackage{graphicx}
\usepackage{xcolor}
\usepackage{array}
\usepackage{tikz-feynman}

\usepackage[normalem]{ulem} 
\usepackage{cancel}

\def\IFIC{Instituto de F\'isica Corpuscular,
Universitat de Val\`encia – Consejo Superior de Investigaciones Cient\'ificas,
Parc Cient\'ific, E-46980 Paterna, Valencia, Spain}

\def\UAB{Grup de F\'isica Te\`orica, Departament de F\'isica, Universitat Aut\`onoma de Barcelona, 08193 Cerdanyola del Vall\`es, (Barcelona)}

\def\ICREA{ICREA, Instituci\'o Catalana de Recerca i Estudis Avan\c{c}ats,\\
Passeig de Llu\'{\i}s Companys 23, 
08010 Barcelona, Spain}

\def\IFAE{Institut de F\'isica d'Altes Energies (IFAE), The Barcelona Institute of Science and Technology, Campus UAB, 08193 Cerdanyola del Vall\`es (Barcelona)}

\title{New Physics in $ b \to s \tau^+\tau^-$ Processes and Correlations with $R(D^{(*)})$ and $B\to K^{(*)}\nu\bar\nu$: An (SM)EFT Analysis}

\author[a,b]{Guillermo Balt\`a,}
\emailAdd{gbalta@ifae.es}
\affiliation[a]{\UAB}
\affiliation[b]{\IFAE}

\author[a,c]{Andreas Crivellin,}
\emailAdd{andreas.crivellin@cern.ch}
\affiliation[c]{\ICREA}

\author[a,b]{Joaquim Matias,}
\emailAdd{matias@ifae.es}

\author[b,d]{and Mart\'in Novoa-Brunet}
\emailAdd{mnovoa@ifae.es}
\affiliation[d]{\IFIC}

 \abstract{
Tauonic flavour-changing neutral-current transitions ($b\to s\tau^+\tau^-$) are sensitive probes of New Physics connected to third-generation quarks and leptons. Motivated by the persistent anomalies in $R(D^{(*)})$ and the indications of enhanced $B^+\to K^{+}\nu\bar\nu$ rates, we revisit and update the predictions for $b\to s\tau^+\tau^-$ processes. Building upon a previous work on a data-driven approach to include the dominant $\psi(2S)$ resonance, we provide predictions for $B\to K^{(*)}\tau^+\tau^-$ ($B_s\to\phi\tau^+\tau^-$) over the full kinematic range. To include possible New Physics contributions within the Weak Effective Theory, we provide semi-analytic expressions for the corresponding branching ratios.

We then perform a comprehensive  dimension-6 SMEFT analysis of New Physics realised at the TeV scale.
 We identify regions of parameter space preferred by current data where the branching ratios of $b\to s\tau^+\tau^-$ processes can be enhanced by several orders of magnitude. 
While current experimental sensitivities remain far above the SM  rates, the enhancements suggested by present flavour data imply that forthcoming limits from LHCb, CMS, and Belle~II will  probe significant regions of the relevant SMEFT parameter space. Furthermore, we highlight that while different scenarios can lead to the same predictions for $B\to K^{(*)}\nu\bar\nu$ branching ratios, they can be disentangled by measuring $B\to K^{(*)}\tau^+\tau^-$, $B_s\to\tau^+\tau^-$ and $B_s\to\phi\tau^+\tau^-$. 
}

\begin{document}
\maketitle
\flushbottom

\section{Introduction}

In the Standard Model (SM) of particle physics, the three generations of fermions are distinguished solely by their Yukawa couplings, whereas the gauge interactions are flavour-universal. Explaining the hierarchy of the Yukawa couplings, i.e., why they are order one for the top quark and $O(10^{-5})$ for the electron, constitutes the flavour puzzle. This suggests that at least the top quark, and probably the whole third generation, is special, due to the approximate $U(2)$ flavour symmetry of the SM. This symmetry implies that searching for New Physics (NP) is most promising in processes involving third-generation fermions~\cite{Barbieri:2011ci,Barbieri:2012uh}, in particular tauonic $B$ meson decays, which involve both third-generation quarks and third-generation leptons~\cite{Calibbi:2015kma,Bordone:2017anc}.

In this context, the concepts of lepton flavour universality (LFU) and LFU violation (LFUV) are particularly important. While after electroweak (EW) symmetry breaking, quark flavour is violated by the CKM matrix, the $W$-lepton-neutrino couplings remain lepton flavour universal.\footnote{This universality is very slightly broken for massive neutrinos. However, the effect is too small to be of any experimental relevance for the observables under consideration.} Note that even though the Yukawa couplings of the leptons are all small, they result in radically different properties for the three generations: the electron is stable, the muon has a finite lifetime, and the tau lepton decays rapidly. However, it is important to keep in mind that the related amplitudes are independent of the lepton masses (disregarding dimension-8 operators and loop effects involving small Yukawa couplings), such that these significant differences between electrons, muons and taus are solely due to kinematics. In this sense, the SM is often said to satisfy LFU. 

Deviations from the SM prediction of LFU  have been actively searched for in precision observables~\cite{Bryman:2021teu}. While several hints were identified in the past~\cite{Crivellin:2021sff} between muons and electrons in $R(K^{(*)})$~\cite{LHCb:2014vgu,LHCb:2017avl,Alguero:2019ptt} and $(g-2)_\mu$~\cite{Aoyama:2020ynm}, after a reanalysis of data and improved theory predictions (in case of $g-2$~\cite{Aliberti:2025beg}) LFU seems to be satisfied now (within errors). In fact, the recent measurements of branching ratios of $B\to K^{(*)}\mu^+\mu^-$ versus $B\to K^{(*)}e^+e^-$~\cite{LHCb:2025ilq}, $B_s\to \phi\mu^+\mu^-$ versus $B_s\to \phi e^+e^-$~\cite{LHCb:2024rto} as well as the difference of angular observables of the muonic versus the electronic mode in $B\to K^{*}\ell^+\ell^-$ measured by LHCb~\cite{LHCb:2025pxz} (so-called $Q_i$~\cite{Capdevila:2016ivx} observables) are consistent with electron-muon universality and point to LFU NP concerning the first and second generation.\footnote{Note that the fading of the hints for LFUV together with the systematic deficit w.r.t.~SM observed in $b\to s \mu^+\mu^-$ transitions~\cite{Descotes-Genon:2012isb,Descotes-Genon:2013vna,LHCb:2013ghj,Descotes-Genon:2013wba,LHCb:2025mqb,LHCb:2014cxe,ATLAS:2018gqc,Belle:2016fev,LHCb:2020gog,CMS:2024atz,LHCb:2021zwz,Gubernari:2022hxn,Gubernari:2020eft}  imply a similar deficit in the corresponding electron channels.}

By contrast, LFU tests in semileptonic charged-current decays involving tau leptons show a different pattern. Measurements of the ratios, defined as
\begin{equation}
R(D^{(*)})=\frac{{\cal B}{(\bar{B} \to D^{(*)} \tau^-\bar{\nu}_\tau)}}{{\cal B}({\bar{B} \to D^{(*)} \ell^-\bar{\nu}_\ell})}\,,
\end{equation}
where $\ell=e,\,\mu$, have consistently lied above the SM prediction since 2012~\cite{BaBar:2012obs,BaBar:2013mob,Belle:2015qfa,Belle:2016ure,Belle:2016dyj,Belle:2017ilt,Belle:2019rba,LHCb:2015gmp,LHCb:2017smo,LHCb:2017rln,LHCb:2023uiv,Belle-II:2025yjp,LHCb:2023zxo,LHCb:2024jll}, when they were first observed by BaBar~\cite{BaBar:2012obs}, pointing towards LFUV related to third-generation tau leptons~\footnote{At the Moriond 2026 conference in La Thuile, preliminary BaBar results were shown which do not indicate an enhancement of these ratios~\cite{BaBar:2026MoriondQCD}.  However, the two BaBar determinations are in substantial tension, particularly for $R(D^*)$, suggesting that the associated systematic uncertainties require further scrutiny~\cite{BaBar:2026MoriondQCD}.}. The analogous $R(J/\psi)$ measurements~\cite{LHCb:2017vlu,LHCb:2026BcJpsiTauSeminar} point in the same direction although with a reduced significance.
It is therefore natural to expect LFUV in the neutral-current decays involving third-generation leptons as well, i.e.~in $b\to s\tau^+\tau^-$ transitions. 

The corresponding branching ratios are predicted to be enhanced by several orders of magnitude by the assumption of NP in purely left-handed operators if one aims to explain the $R(D^{(*)})$ anomalies~\cite{Alonso:2015sja,Crivellin:2017zlb,Calibbi:2017qbu,Capdevila:2017iqn}.
Such an enhancement is in agreement with NP interpretations of the anomalies in $b\to s\ell^+\ell^-$ global fits~\cite{Altmannshofer:2021qrr,Alguero:2022wkd,Hurth:2025vfx}\footnote{{Global fits to $b\to s\ell^+\ell^-$ data are sensitive to the local form factors and to the treatment of non-local hadronic matrix elements, both of which remain under active study~\cite{LHCb:2023gpo,LHCb:2026enw,Capdevila:2025drq,Ciuchini:2026hxy, Isidori:2025dkp}.
Nevertheless, it was recently pointed out in Ref.~\cite{Alvarez-Cartelle:2026nhp} that, with additional data on inclusive decays, taking into account the constraints they impose on the corresponding exclusive modes could make it possible to disentangle NP from a hadronic explanation.}}. 
It can be explained by the same operator via a tau loop~\cite{Bobeth:2011st,Crivellin:2018yvo} and is consistent with the current upper experimental limits (90\% CL)~\cite{LHCb:2017myy,Belle-II:2026ism,LHCb:2025lcw,Belle-II:2025lwo} from LHCb, Belle (and Belle II) and BaBar
\begin{equation}
\begin{aligned}
    \mathcal{B}\left(B_s\to \tau^+\tau^-\right)<5.2\times 10^{-3}\,,
\\
    \mathcal{B}\left(B^+\to K^+\tau^+\tau^-\right)<5.6\times 10^{-4}\,, 
\\
    \mathcal{B}\left(B^0\to K^{0*}\tau^+\tau^-\right)<2.5\times 10^{-4}\,,
\\
    \mathcal{B}\left(B_s\to\phi\tau^+\tau^-\right)<4.1\times 10^{-4}\,.
    \end{aligned}
\end{equation}
Furthermore, the indication for an enhancement of $\mathcal{B}(B\to K\nu\bar{\nu})$~\cite{Belle-II:2023esi} is likely related to third-generation lepton neutrinos~\cite{Allwicher:2023xba}, if one disregards the possibility of light NP~\cite{Altmannshofer:2023hkn,Bolton:2024egx,Bolton:2025fsq,Bolton:2025lnb}, further supporting the hypothesis of NP related to tau leptons~\footnote{The present pattern of an enhanced $B\to K\nu\bar\nu$ rate together with the $B\to K^*\nu\bar\nu$ bound favours a right-handed $b\to s$ quark current coupled to left-handed lepton doublets~\cite{Allwicher:2023xba}. In the SMEFT, unlike the left-handed quark-current contribution, no cancellation between singlet and triplet operators is possible, so the same coefficient necessarily contributes to $b\to s\ell^+\ell^-$. The strong constraints on the electron and muon channels  of $\mathcal{O}(10\%)$ therefore favour a tau-dominated coupling.}.

While $B_s\to\tau^+\tau^-$ is theoretically clean, its branching ratio is chirally suppressed and experimentally challenging at hadron colliders due to the missing energy involved, which hinders reconstruction. Furthermore, it can only be measured at LHCb and CMS but not at Belle II (at the $\Upsilon(4S)$ resonance), while $B\to K^{(*)}\tau^+\tau^-$ and $B_s\to \phi\tau^+\tau^-$ allow for a better reconstruction of the events since they result in more tracks in the detector, but these semileptonic decays receive long-distance contributions from hadronic resonances. In a clean lepton-collider environment, one can cut on the invariant mass of the tau pair to remove the dominant $\psi(2S)$ resonance and the associated uncertainties, as tagging of the second $B$ meson is possible. Notice that such a $q^2$ cut reduces the branching ratio, even in the case of large NP, by a factor $\approx 1.5$~\cite{Balta:2026yea}, making an observation even more challenging. At hadron colliders, the corresponding reconstruction requires additional kinematic assumptions and would lead to limited $q^2$-resolution.  Therefore, we calculated in Ref.~\cite{Balta:2026yea} the branching ratios for $B\to K^{(*)}\tau^+\tau^-$ over the full kinematically accessible $q^2$ range, taking into account the effect of the hadronic resonances, in particular the dominant $\psi(2S)$ one. For this, we used both a full data-driven approach and a simplified approach applying the narrow-width approximation, finding good agreement between the two methods.

In this paper, building on the result of  Ref.~\cite{Balta:2026yea}, we consider $b\to s\tau^+\tau^-$ decays within the weak effective theory and derive their branching ratios in terms of the Wilson coefficients $C_{9(10)}^\tau$
including the chirally flipped $C_{9'(10')}^\tau$. Next, in Section~\ref{sec:SMEFT}, we study the 
SM Effective Field Theory (SMEFT) with manifest $SU(2)_L$ invariance and discuss the impact of $R(D^{(*)})$ and $\mathcal{B}(B\to K^{(*)}{\nu\bar{\nu}})$ by expressing the $b\to s\tau^+\tau^-$ rates in terms of these observables. We conclude in Section~{\ref{sec:conclusions}}.

\section{$b \to s \tau^+\tau^-$ Processes in the Weak Effective Theory}
\label{sec:WET}

\subsection{SM predictions and inclusion of hadronic resonances}\label{subsec:SMpredictions}

Let us first recall some experimental considerations concerning the $q^2$ range over which the branching ratios of the semileptonic $b\to s\tau^+\tau^-$ processes are integrated. Belle II can measure the dilepton invariant mass via the momentum of the decay meson $K$ or $K^*$, since it can reconstruct the $B$ momentum through tagging of the second entangled $B$ meson. Instead, in a hadronic environment such as LHCb or CMS, no equivalent tagging of the companion $B$ hadron is possible. Therefore, measuring the $K^{(*)}$ momentum is not sufficient to reconstruct the tau-pair invariant mass, and any approximation to infer the dilepton invariant mass requires additional kinematic assumptions and will suffer from resolution issues. As a consequence, full $q^2$-range measurements starting at $4m_{\tau}^2$ at LHCb and CMS are experimentally better motivated, although they include the dominant $\psi(2S)$ resonance. Corresponding predictions for the full range are thus required.

In Ref.~\cite{Balta:2026yea}, we used a data-driven approach to calculate the branching ratios for the decays $B\to K^{(*)}\tau^+\tau^-$, integrated over the full kinematically accessible $q^2$ range. Here, we will distinguish two kinematic regions: the full kinematically accessible phase space, from $q^2_{min}=4 m_\tau^2$ to $q^2_{i,{\rm max}}=22, 19, 18.8$\,GeV$^2$ for $i=K^+,K^*,\phi$, and  from $q^2_{\rm min}=14.18$\,GeV$^2$ to $q^2_{i,{\rm max}}$ (as used by Belle II). Integrating over these intervals, we obtain~\cite{Balta:2026yea}
\begin{equation}
\label{SM predictions}
\begin{array}{rcl}
{\cal B}(B^+\to K^+\tau^+\tau^-)_{\rm SM}^{[4 m_{\tau}^2,q_{K,{\rm max}}^2]}&=& 1.86\substack{+0.17\\-0.16}\times 10^{-6}\ ,\\[1ex]
{\cal B}(B\to K^*\tau^+\tau^-)_{\rm SM}^{[4 m_{\tau}^2,q_{K^*,{\rm max}}^2]}&=& 1.50\substack{+0.23\\-0.22} \times 10^{-6}\ ,\\[1ex]
{\cal B}(B^+\to K^+\tau^+\tau^-)_{\rm SM}^{[14.18,q_{K,{\rm max}}]}&=& 1.54 \, \substack{+0.16 \\ -0.13}\times 10^{-7}\ ,\\[1ex]
{\cal B}(B\to K^*\tau^+\tau^-)_{\rm SM}^{[14.18,q_{K^*,{\rm max}}]}&=& 1.36 \, \substack{+0.61 \\ -0.40} \times 10^{-7}\ .
\end{array}
\end{equation}

Furthermore, in Ref.~\cite{Balta:2026yea}, we compared the fully data-driven approach to the narrow-width approximation and found good agreement. This allows us to provide in this work a first estimate of $\mathcal{B}(B_s \to \phi\tau^+\tau^-)$ over the full phase space, by applying the narrow-width approximation, since the information needed for an analogous data-driven determination is not yet available. As a result, we find

\begin{gather} 
{\cal B}(B_s\to \phi\tau^+\tau^-)_{\rm SM}^{[4 m_{\tau}^2,q_{\phi,{\rm max}}^2]}= 1.7\substack{+0.2\\-0.2}\times 10^{-6}\,.
\end{gather}

\subsection{New Physics sensitivity}\label{sec:sensitivity}

We now calculate the dependence of the branching ratios on the NP Wilson coefficients in the Weak Effective Theory (WET).\footnote{Note that since the operators under consideration correspond to conserved currents under QCD, the running below the EW scale is only due to QED effects and thus small.} At the $B$ meson scale, the effective Hamiltonian (following the conventions of e.g.~Ref.~\cite{Buras:1994dj}) 
\begin{equation}
{\cal H}_{\rm eff}=- \frac{4 G_F}{\sqrt{2}} V_{tb}V_{ts}^* \sum C_i {\cal O}_i\,, \label{bstautauH}
\end{equation}
 with 
\begin{equation}
\begin{aligned}
{\cal O}_{9}^\tau&=\frac{\alpha}{4\pi}[\bar{s}_L \gamma^\mu b_L][\bar{\tau}\gamma_\mu \tau]\,,
\quad \quad 
{\cal O}_{10}^\tau=\frac{\alpha}{4\pi}[\bar{s}_L \gamma^\mu b_L][\bar{\tau}\gamma_\mu\gamma_5 \tau]\,, \nonumber \cr
{\cal O}_{9'}^\tau&=\frac{\alpha}{4\pi}[\bar{s}_R \gamma^\mu b_R][\bar{\tau}\gamma_\mu \tau]\,, 
\quad \quad 
{\cal O}_{10'}^\tau=\frac{\alpha}{4\pi}[\bar{s}_R \gamma^\mu b_R][\bar{\tau}\gamma_\mu\gamma_5 \tau]\,,
\end{aligned}
\end{equation}
describes $b\to s \tau^+\tau^-$ transitions. We disregard scalar and tensor operators. Note that scalar operators would lead to a dominant effect in $B_s\to \tau^+\tau^-$, and tensor operators affecting $b\to s \tau^+\tau^-$ are anyway not generated in the SMEFT at the dimension-6 level~\cite{Alonso:2014csa}. Additionally, all NP Wilson coefficients are assumed to be real.

We now provide semi-analytic formulas for the branching ratios (integrated over the full kinematically accessible region) as a function of the  NP Wilson coefficients $C_9^{\tau\, \rm NP}$, $C_{10}^{\tau\, \rm NP}$ and their primed counterparts:
\begin{align}
10^8\times{\rm Br}(B_{(s)}\to X\tau^+\tau^-)&= {\rm SM}^X+\alpha_1^X C_{9}^{\tau\, \rm NP}
+ \alpha_2^X C_{10}^{\tau\, \rm NP} + \alpha_3^X {(C_{9}^{\tau\, \rm NP})}^2 + \alpha_4^X  {(C_{10}^{\tau\, \rm NP})}^2 \nonumber \\
&+
\beta_1^X C_{9'}^{\tau\, \rm NP}
+ \beta_2^X C_{10'}^{\tau\, \rm NP} + \beta_3^X  {(C_{9'}^{\tau\, \rm NP})}^2 + \beta_4^X  {(C_{10'}^{\tau\, \rm NP})}^2  \label{eq:Brsemianalytic}\\
&+
\gamma_1^X C_{9}^{\tau\, \rm NP} C_{9'}^{\tau\, \rm NP} + \gamma_2^X  {C_{10}^{\tau\, \rm NP}} {C_{10'}^{\tau\, \rm NP}}\,.
\nonumber 
\end{align}

\begin{table}[t]
\centering
\begin{tabular}{c|ccccccccccc}
\hline
$X$ & $\mathrm{SM}^X$ & $\alpha_1^X$ & $\alpha_2^X$ & $\alpha_3^X$ & $\alpha_4^X$ & $\beta_1^X$ & $\beta_2^X$ & $\beta_3^X$ & $\beta_4^X$ & $\gamma_1^X$ & $\gamma_2^X$ \\
\hline
$K$ & 185 & 2.1 & -5.4 & 0.31 & 0.62 & 2.1 & -5.4 & 0.31 & 0.62 & 0.63 & 1.2 \\
$K^*$ & 148 & 4.9 & -1.7 & 0.63 & 0.20 & -3.8 & 1.5 & 0.63 & 0.20 & -0.86 & -0.35 \\
$\phi$ & 170 & 4.8 & -1.9 & 0.71 & 0.22 & -3.6 & 1.7 & 0.71 & 0.22 & -1.1 & -0.40 \\
\hline
\end{tabular}
\caption{Constants parametrising the branching ratios in Eq.~\eqref{eq:Brsemianalytic} for the case $B^+\to K^+\tau^+\tau^-$ ($X=K$), $B^0\to K^{*0}\tau^+\tau^-$ ($X=K^*$), and $B_s \to \phi \tau^+\tau^-$ ($X=\phi$), integrated over the full kinematically accessible region. }
\label{tab:parameters}
\end{table}

Here $C_i^{\tau\,{\rm NP}}$ denotes the NP contribution to the corresponding WET coefficient, $C_i^\tau=C_i^{\tau\,{\rm SM}}+C_i^{\tau\,{\rm NP}}$. For each final state $X=K,K^*,\phi$, $\mathrm{SM}^X$ corresponds to the full range SM branching ratio quoted in subsection~\ref{subsec:SMpredictions}. The terms linear in the NP coefficients describe SM-NP interference, whereas the quadratic terms give the pure NP contribution, being the dominant terms for large NP. The numerical values of $\mathrm{SM}^X$ and of the coefficients $\alpha_i^X$, $\beta_i^X$, and $\gamma_i^X$ are collected in Table~\ref{tab:parameters}.

\begin{figure}[htbp]
    \centering
    \begin{subfigure}[b]{0.48\textwidth}
        \centering
        \includegraphics[width=\textwidth]{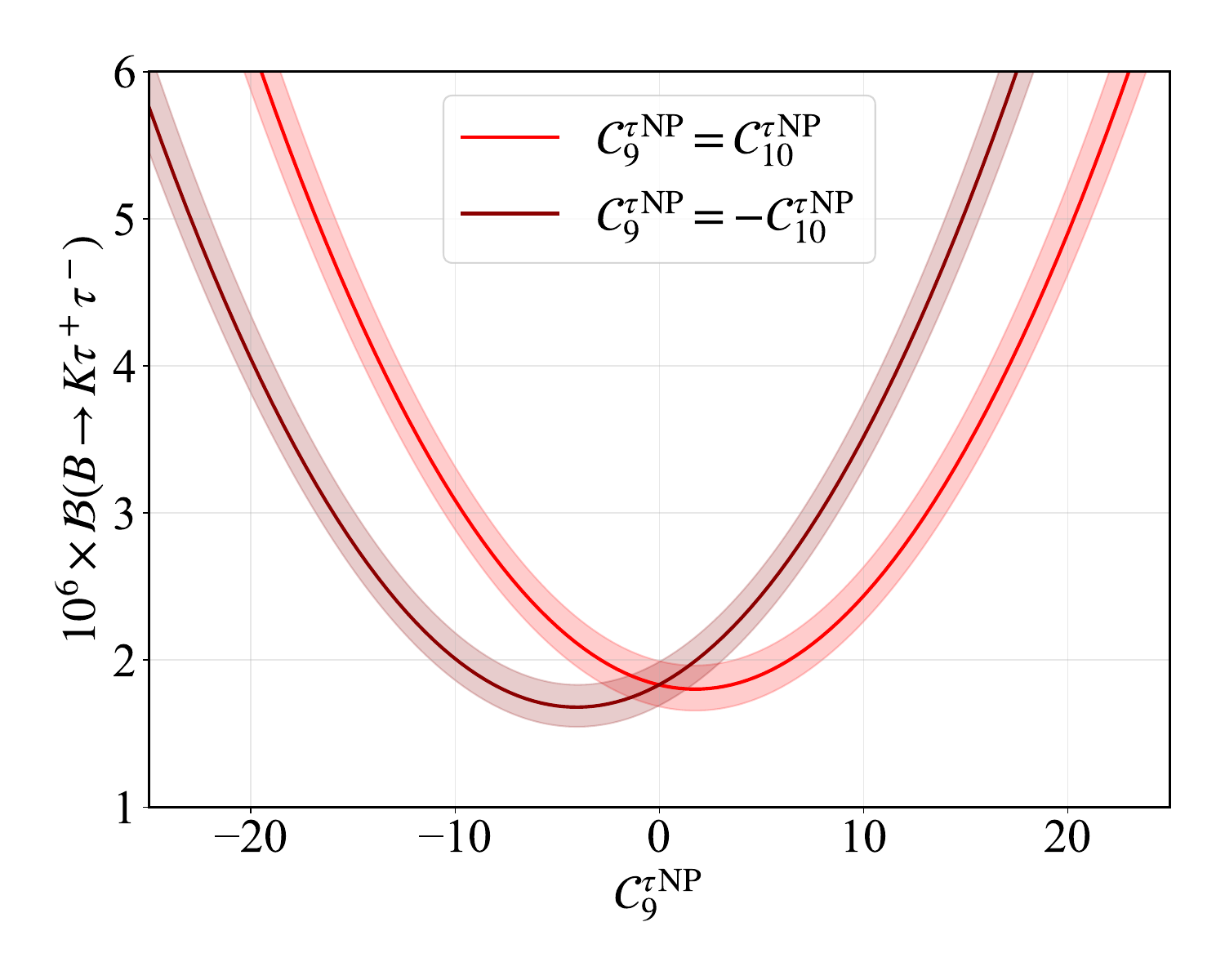}
        \caption{$B\to K\tau^+\tau^-$.}
        \label{fig:BKtautau vs C9}
    \end{subfigure}
    \hfill 
    \begin{subfigure}[b]{0.48\textwidth}
        \centering
        \includegraphics[width=\textwidth]{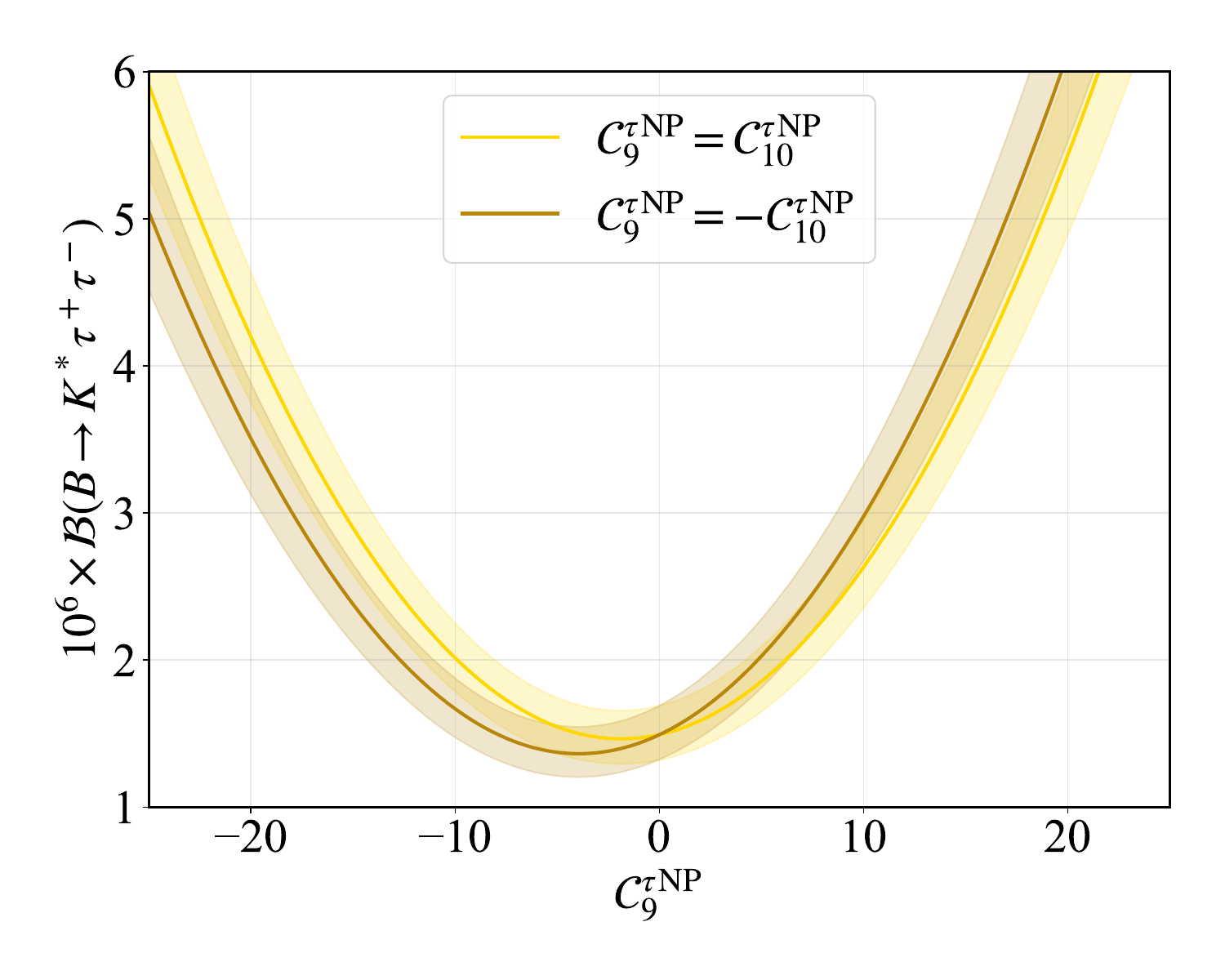}
        \caption{$B\to K^*\tau^+\tau^-$.}
        \label{fig:BKstartautau vs C9}
    \end{subfigure}
    \caption{Branching ratios of $B\to K \tau^+\tau^-$ (left) and  $B\to K^* \tau^+\tau^-$ (right)  as a function of $C_{9}^{\tau\, \rm NP}$. The case where $C_{9}^{\tau\, \rm NP}=(-)C_{10}^{\tau\, \rm NP}$ is shown in light (dark) colour bands. The width of the bands indicates the theoretical uncertainty.}
    \label{fig:BRs vs C9}
\end{figure}

The expressions above give the central values only. Fig.~\ref{fig:BRs vs C9} shows the corresponding theory uncertainties for the benchmark relations $C_{9}^{\tau\, \rm NP}=\pm C_{10}^{\tau\, \rm NP}$, as motivated by models addressing $R(D^{(*)})$ and/or generating $C_9^\text{U}\equiv C^{e\;{\rm NP}}_9=C^{\mu\;{\rm NP}}_9$~\footnote{In particular, the purely left-handed NP effect ($C_{9}^{\tau\, \rm NP}=-C_{10}^{\tau\, \rm NP}$) is motivated by the $S_1+S_3$ and $U_1$ LQ models, while a left-handed current on the quark side and a right-handed current on the lepton side ($C_{9}^{\tau\, \rm NP}=+C_{10}^{\tau\, \rm NP}$) can be generated by the $R_2$ LQ model~\cite{Crivellin:2022mff}. Note that while both cases can explain $C_9^\text{U}$ via an off-shell photon penguin, only the former case has the potential to account for $R(D^{(*)})$. We will see this in the next section, where we use SMEFT to relate $b\to s\tau^+\tau^-$ processes to $b\to c\tau\nu$ and $b\to s\nu\bar{\nu}$.}.
For the purpose of this study, from now on we will restrict ourselves to purely left-handed lepton NP currents ($C_{9}^{\tau\, \rm NP}=-C_{10}^{\tau\, \rm NP}$) as motivated by current tensions in $b\to c\tau\nu$ observables. If these tensions are reduced  and no longer point towards NP, right-handed lepton NP currents ($C_{9}^{\tau\, \rm NP}=+C_{10}^{\tau\, \rm NP}$) would instead become a natural possibility to account for $C_9^\text{U}$.

\section{SMEFT analysis and correlations with $b\rightarrow s\nu\bar{\nu}$ and $b\rightarrow c\tau\nu$}
\label{sec:SMEFT}

The SMEFT parametrises NP realised above the EW scale via higher-dimensional operators. At the dimension-6 level, the extra terms allowed in the Lagrangian are generically written as
\begin{equation}
\label{6dim SMEFT}
\mathcal L^{(6)}_\mathrm{SMEFT} \supset  \sum_i \frac{\Tilde{C}_i}{\Lambda^2} {\mathcal O}_i\,,
\end{equation}
manifestly respecting $SU(2)_L$ gauge invariance. This, for example, leads to relations between charged and neutral current semileptonic operators involving left-handed fields. From now on, we will work with dimensionful Wilson coefficients defined as $C_i\equiv\Tilde{C}_i/\Lambda^2$. 

In this framework, our previous analysis~\cite{Capdevila:2017iqn} showed that the anomalies in $R(D^{(*)})$ predict, under quite general assumptions, an enhancement of $b\to s\tau^+\tau^-$ processes by several orders of magnitude. To be more specific, we assumed that 
\begin{itemize}
    \item NP is realised at the scale $\Lambda$ (of the order of a few TeV) such that one can expand in $v/\Lambda$ and use the leading order of the SMEFT in this expansion (corresponding to dim-6 operators).
    \item The NP effect is related to tau leptons.
    \item NP only enters a minimal set of operators with left-handed vector currents
        \begin{eqnarray}
\big{[}\mathcal{O}_{\ell q}^{(1)}\big{]}_{ijkl}&=&[{\bar Q}_i \gamma_\mu Q_j] [{\bar L}_k \gamma^\mu L_l]  \nonumber \cr
\big{[}\mathcal{O}_{\ell q}^{(3)}\big{]}_{ijkl}&=&[{\bar Q}_i \gamma_\mu \sigma^I Q_j] [{\bar L}_k \gamma^\mu \sigma^I L_l]\end{eqnarray}   
    necessary to explain $R(D^{(*)})$ without violating constraints from $B\to K^{(*)}\nu\Bar{\nu}$.\footnote{These operators are generated most naturally either in the $U_1$ vector-leptoquark model~\cite{Calibbi:2015kma,Barbieri:2016las,DiLuzio:2017vat,Calibbi:2017qbu,Bordone:2017bld,Blanke:2018sro,Crivellin:2018yvo} or in the $S_1+S_3$ scalar leptoquark model~\cite{Crivellin:2017zlb,Crivellin:2019dwb,Gherardi:2020qhc}. While the former predicts $C^{(1)}_{2333}=C^{(3)}_{2333}$ at tree-level, the latter needs tuning for this case and thus favours larger effects in $B\to K^{(*)}\nu \bar \nu$. The relation between $b\to s\nu\bar\nu$ and $R(D^{(*)})$ in leptoquark models has been studied e.g.~in Refs.~\cite{Fajfer:2018bfj,Descotes-Genon:2020buf,Allwicher:2023xba,Chen:2023wpb,Rosauro-Alcaraz:2024mvx,Chen:2025npb,Crivellin:2025qsq}.} 
    \item NP has a generic (not hierarchical) flavour structure such that $\big{[}\mathcal{O}_{\ell q}^{(1)}\big{]}_{2333}$ and $\big{[}\mathcal{O}_{\ell q}^{(3)}\big{]}_{2333}$ are the relevant operators involved and the CKM-suppressed effect of the operator $\big{[}\mathcal{O}_{\ell q}^{(3)}\big{]}_{3333}$ in the $b\to c\tau\nu$ current can be neglected.
    \item There is no (tree-level) NP effect in the neutrino mode $\big(\big{[}C_{\ell q}^{(1)}\big{]}_{2333}=\big{[}C_{\ell q}^{(3)}\big{]}_{2333}\big)$.
\end{itemize}
As a consequence, the large contribution required to explain the anomaly in $R(D^{(*)})$ of $\approx10$\% w.r.t.~the tree-level SM contribution, has a huge impact on the loop-induced $b\to s \tau^+\tau^-$ processes such that decays like $B_s\to \tau^+\tau^- $, $B\to K^{(*)}\tau^+\tau^-$, and $B_s\to\phi\tau^+\tau^-$ can have branching ratios up to $10^{-5}$. 

In recent years, an enhanced decay rate of $B\to K\nu\bar{\nu}$ was observed by Belle II~\cite{Belle-II:2023esi}. Combining the available measurements~\cite{Belle-II:2021rof,Belle:2013tnz,BaBar:2013npw,BaBar:2010oqg,Belle:2017oht}, one obtains
\begin{equation}
    \mathcal{B}\left(B^+\rightarrow K^+\nu\bar{\nu}\right)=(1.3\pm0.4)\times 10^{-5}\,,
\end{equation}
which is significantly higher than the corresponding SM prediction~\cite{Becirevic:2023aov,Parrott:2022zte} of\footnote{Note that this value does not include the tree-level contribution~\cite{Kamenik:2009kc} which is subtracted in the experimental analysis~\cite{Belle-II:2023esi}.}
\begin{equation}
\mathcal{B}_\text{SM}\left(B^+\rightarrow K^+\nu\bar{\nu}\right)=(4.44\pm 0.14\pm0.27)\times 10^{-6} \,.
\end{equation}
This experimental information suggests relaxing the assumption $\big{[}C_{\ell q}^{(1)}\big{]}_{2333}=\big{[}C_{\ell q}^{(3)}\big{]}_{2333}$ to allow for an enhancement of $b\to s\nu\bar{\nu}$ transitions at tree-level. Furthermore, for $B\to K^*\nu\bar{\nu}$ only an upper bound exists, which does not point towards an enhancement~\cite{Belle:2017oht},
\begin{equation}
    \mathcal{B}\left(B^0\rightarrow K^{*0}\nu\bar{\nu}\right)<2.7\times 10^{-5}\,\,\, (90\%\,\, \rm CL)\,,
\end{equation}
w.r.t.~the SM prediction~\cite{Becirevic:2023aov},\footnote{In fact, as can be seen from Eq.~\eqref{eq:enhancement}, at 90\% CL, the limit on $B^0\rightarrow K^{*0}\nu\bar{\nu}$ is slightly stronger than the central value for $B^+\rightarrow K^{+}\nu\bar{\nu}$, once normalised to the SM prediction.}
\begin{equation}
\mathcal{B}_{\rm SM}(B^0\to K^{0\ast}\nu\bar{\nu}) = (9.05\pm 1.37) \times 10^{-6}\,.
\end{equation} 
This could require the presence of right-handed $b\to s$ currents~\cite{Allwicher:2023xba}. 

Therefore, this SMEFT analysis has three main goals. First, we employ the latest theoretical predictions over the entire $q^2$ range, incorporating both the $\psi(2S)$ resonance contribution and updated theory inputs. Second, we assess the role of the recent $b\to s\nu\bar{\nu}$ measurements. In contrast to the tree-level induced observables $R(D^{(*)})$, whose effects can be enhanced by orders of magnitude relative to the SM, the impact of $b\to s\nu\bar{\nu}$ observables is expected to be more modest, at the level of $\mathcal{O}(1)$, since these transitions are loop-induced in the SM, similarly to $b\to s\tau^+\tau^-$. Third, we explore the implications of a hierarchical flavour structure motivated by less minimal flavour violation~\cite{Barbieri:1995uv,Barbieri:1997tu,Barbieri:2012bh}\footnote{Similarly, ``standard" minimal flavour violation~\cite{Chivukula:1987fw,Hall:1990ac,Buras:2000dm} (MFV) is based on $U(3)^3$~\cite{DAmbrosio:2002vsn}. However, $U(3)^3$ is strongly broken to $U(2)^3$ by the large third-generation Yukawa couplings.}, whereby the Wilson coefficient $\big{[}C_{\ell q}^{(3)}\big{]}_{3333}$ can, despite CKM suppression, substantially modify the correlations between charged- and neutral-current observables~\cite{Fuentes-Martin:2019mun,Marzocca:2024hua}.

\subsection{WET and Observables}

We provided in the previous section the expressions for $b\to s\tau^+\tau^-$ transitions in the WET. Here, we want to give the analogous expressions for the other observables of interest in this paper.

\subsubsection{$B\to K^{(*)}\nu\bar{\nu}$}
\label{BtoKnunu}

For $B\to K^{(*)}\nu\bar{\nu}$ it is convenient to define
\begin{equation}
R_{K^{(*)}}^{\nu}\equiv \frac{\mathcal{B}(B\rightarrow K^{(*)}\nu\bar{\nu})}{\mathcal{B}_{\rm SM}(B\rightarrow K^{(*)}\nu\bar{\nu})}\,,
\end{equation}
which at the moment are measured to be~\cite{Belle-II:2023esi,Marzocca:2024hua,Belle:2017oht}
\begin{equation} \label{eq:enhancement}
\begin{aligned}
  R_K^{\nu , \rm exp}&=2.93\pm0.90\,, \\
  R_{K^*}^{\nu , \rm exp}&<2.7\,\, \text{(90$\%$ CL)}\,.
\end{aligned}
\end{equation}

In the low-energy effective theory describing $b\to s\nu\bar{\nu}$ decays, the effective Lagrangian~\cite{Altmannshofer:2009ma,Buras:2014fpa,Becirevic:2023aov} is
\begin{align}
\label{eq:eft-bsnunu}
\mathcal{L}_\mathrm{eff}^{b\to s\nu\bar{\nu}} =  \dfrac{4 G_F}{\sqrt{2}} V_{tb} V_{ts}^* \sum_a C_a\, \mathcal{O}_a+\mathrm{h.c.}\,, 
\end{align}
with
\begin{align}
\label{eq:eft-ops}
\mathcal{O}_{L}^{\nu_i\nu_j} &=\dfrac{e^2}{(4\pi)^2}(\bar{s}_L \gamma_\mu b_L)(\bar{\nu}_i \gamma^\mu (1-\gamma_5)\nu_j)\,,\nonumber\\[0.3em]
\mathcal{O}_R^{\nu_i\nu_j} &=\dfrac{e^2}{(4\pi)^2}(\bar{s}_R \gamma_\mu b_R)(\bar{\nu}_i \gamma^\mu (1-\gamma_5)\nu_j)\,.
\end{align}
We parametrise the NP contributions as
\begin{align}
R_{K^{(*)}}^{\nu} &= 1+\delta \mathcal{B}_{K^{(\ast)}}^{\nu}\,,
\end{align}
so that the NP piece, $\delta \mathcal{B}_{K^{(\ast)}}^{\nu}$, can be expressed in terms of $\delta C_{L,R}^{\nu_i\nu_j}$. If we write $C_{L,R}^{\nu_i\nu_j}=  \delta_{ij} C_{L,R}^\mathrm{SM} + \delta C_{L,R}^{\nu_i\nu_j}$, then we have~\cite{Buras:2014fpa,Becirevic:2023aov}: 
\begin{align}
\label{NP in bsnunu}
\begin{split}
\delta \mathcal{B}_{K^{(\ast)}}^{\nu} &=  \sum_{i}\dfrac{2\mathrm{Re}[C_L^\mathrm{SM}\,(\delta C_{L}^{\nu_i\nu_i}+\delta C_{R}^{\nu_i\nu_i})]}{3|C_{L}^\mathrm{SM}|^2}\\
&+\sum_{i,j}\dfrac{|\delta C_{L}^{\nu_i\nu_j}+\delta C_{R}^{\nu_i\nu_j}|^2}{3|C_L^\mathrm{SM}|^2}- \eta_{K^{(\ast)}}\sum_{i,j} \dfrac{\mathrm{Re}[\delta C_R^{\nu_i\nu_j}(C_{L}^\mathrm{SM}\delta_{ij}+\delta C_{L}^{\nu_i\nu_j})]}{3|C_{L}^\mathrm{SM}|^2}\,,
\end{split}
\end{align}
where the sum over neutrino flavour indices is understood, $i,j \in \lbrace 1,2,3 \rbrace$. In the above expression, {$C_L^\mathrm{SM}=-6.32$, $C_R^\mathrm{SM}=0$}, $\eta_K=0$, and $\eta_{K^\ast}=3.33(7)$~\cite{Buras:2014fpa}. Considering NP only in the third generation, $\delta \mathcal{B}_{K^{(\ast)}}^{\nu}$ in Eq.~\eqref{NP in bsnunu} simplifies to
\begin{align}
\label{NP in bsnunu 3rd gen}
\begin{split}
\delta \mathcal{B}_{K^{(\ast)}}^{\nu_\tau} &=  \dfrac{2\mathrm{Re}[C_L^\mathrm{SM}\,(\delta C_{L}^{\nu_\tau}+\delta C_{R}^{\nu_\tau})]}{3|C_{L}^\mathrm{SM}|^2}\\
&+\dfrac{|\delta C_{L}^{\nu_\tau}+\delta C_{R}^{\nu_\tau}|^2}{3|C_L^\mathrm{SM}|^2}- \eta_{K^{(\ast)}} \dfrac{\mathrm{Re}[\delta C_R^{\nu_\tau}(C_{L}^\mathrm{SM}+\delta C_{L}^{\nu_\tau})]}{3|C_{L}^\mathrm{SM}|^2}\,.
\end{split}
\end{align}

\subsubsection{$b\to c\tau\nu$}
Concerning the current experimental and theoretical status of the $b\to c\tau\nu$ transitions, the world averages for the ratios $R(D)$, $R(D^*)$, and $R(J/\psi)$ obtained from measurements by BaBar, Belle, Belle II, and LHCb~\cite{BaBar:2013mob,Belle:2015qfa,Belle:2016ure,Belle:2016dyj,Belle:2017ilt,Belle:2019rba,LHCb:2015gmp,LHCb:2017smo,LHCb:2017rln,LHCb:2023uiv,Belle-II:2025yjp,LHCb:2023zxo,LHCb:2024jll,LHCb:2017vlu,LHCb:2026BcJpsiTauSeminar} are
\begin{gather}
    R(D)=0.358\pm0.024  \,,\\
    R(D^*)=0.281\pm0.011 \,,\\
    R(J/\psi)=0.51\pm0.14\,.
\end{gather}
The corresponding SM predictions for
 $b\rightarrow c\tau\nu$ modes are
\begin{equation}
\begin{aligned}
    R(D)_\text{SM}=0.296\pm 0.004 \;\;(0.2938\pm 0.0054) \,,\\
    R(D^*)_\text{SM}=0.254\pm 0.005 \;\;(0.2582\pm 0.0051)\,,
\end{aligned}
\end{equation}
from HFLAV~\cite{HeavyFlavorAveragingGroupHFLAV:2024ctg} (FLAG~\cite{FlavourLatticeAveragingGroupFLAG:2024oxs})
and 
\begin{equation}
R(J/\psi)_\text{SM}=0.2597\pm 0.0027\,,
\end{equation}
from HPQCD~\cite{Harrison:2025yan}.
Defining the effective Lagrangian as
\begin{eqnarray}
\label{eq:leff1} 
{\cal L}_{\rm eff} 
=
- \frac{4G_F V_{cb}}{\sqrt{2}}
\Big(1 + \epsilon_L \Big) \bar{\tau}  \gamma_\mu  P_L \nu_{\tau} \cdot \bar{c}   \gamma^\mu P_L b+{\rm h.c.}
\end{eqnarray}
the relative effect w.r.t.~the SM in tauonic $B$ decays is given by
\begin{equation}
    \frac{R(D^{(*)})}{R(D^{(*)})_\text{SM}}= \frac{R(J/\psi)}{R(J/\psi)_\text{SM}}=\left|1+\epsilon_L\right|^2\,.
\end{equation}

\subsubsection{$b\to s\ell^+\ell^-$}

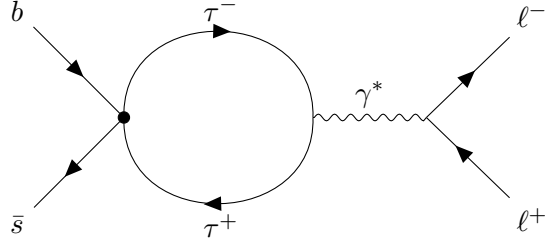
\begin{figure}[h] 
    \centering
    \begin{tikzpicture}
        \begin{feynman}
            \vertex [dot] (origin) {}; 
            \vertex [above left=2cm of origin] (b_in) {$b$};
            \vertex [below left=2cm of origin] (s_in) {$\bar{s}$};
            \vertex [right=2.5cm of origin] (v_photon);
            \vertex [right=1.5cm of v_photon] (v_decay);
            \vertex [above right=1.5cm of v_decay] (l_minus) {$\ell^-$};
            \vertex [below right=1.5cm of v_decay] (l_plus) {$\ell^+$};
            \diagram* {
                (b_in) -- [fermion] (origin),
                (s_in) -- [anti fermion] (origin),
                (origin) -- [fermion, half left, edge label=$\tau^-$] (v_photon),    
                (origin) -- [anti fermion, half right, edge label'=$\tau^+$] (v_photon),
                (v_photon) -- [photon, edge label=$\gamma^*$] (v_decay),
                (v_decay) -- [fermion] (l_minus),
                (v_decay) -- [anti fermion] (l_plus),
            };
        \end{feynman}
    \end{tikzpicture}
    \caption{The tau-loop off-shell photon penguin with the photon coupling to $\ell^+ \ell^-$, generating a mixing of ${\cal O}_{9}^\tau$ into ${\cal O}_9^{\rm U}$.}
    \label{fig: Feynman diagram tau loop}
\end{figure}

For light leptons (i.e.~$\ell=e,\mu$), we use the same conventions as for the $\tau$ case (with obvious replacements). Nevertheless, a large  $C_9 ^{\tau\, \rm NP}$ Wilson coefficient has important loop effects relevant for $b\to s\ell^+\ell^-$ transitions with light leptons. For this, one should also consider the off-shell photon penguin~\cite{Crivellin:2018yvo}, shown in Fig.~\ref{fig: Feynman diagram tau loop}, leading to a mixing of an $\bar s b \bar \tau \tau$ operator into an $\bar s b \bar \ell \ell$ operator giving rise to the relation~\cite{Alguero:2022wkd}
\begin{equation} \label{eq: C9U}
    C_9^\text{U}=-\frac{\alpha}{3\pi}\log\left(\frac{\Lambda^2}{\mu_b^2}\right)C_9^{\tau\, \rm NP}\,,
\end{equation}
where we identify $C_9^\text{U}\equiv C_9^{\mu\, \rm NP}=C_9^{e\, \rm NP}$ (see subsection~\ref{sec:sensitivity}).

\subsection{Matching of SMEFT onto WET}

Within the SMEFT Lagrangian at the dimension-6 level given in Eq.~\eqref{6dim SMEFT}, we consider the four-fermion operators
\begin{align}\label{eq:4f}
\big{[}\mathcal{O}_{\ell q}^{(1)}\big{]}_{2333} &=\big{[}\Bar{Q}_2 \gamma_\mu Q_3\big{]}\big{[}\Bar{L}_3\gamma^\mu L_3\big{]}\,,\\[0.4em]
\big{[}\mathcal{O}_{\ell q}^{(3)}\big{]}_{2333} &=\big{[}\Bar{Q}_2 \gamma_\mu \sigma^I Q_3\big{]}\big{[}\Bar{L}_3\gamma^\mu \sigma^I L_3\big{]}\,,\\[0.4em]
\big{[}\mathcal{O}_{\ell q}^{(3)}\big{]}_{3333} &=\big{[}\Bar{Q}_3 \gamma_\mu \sigma^I Q_3\big{]}\big{[}\Bar{L}_3\gamma^\mu \sigma^I L_3\big{]}\,,\\[0.4em]
\big{[}\mathcal{O}_{\ell d}\big{]}_{2333} &=\big{[}\Bar{d}_{2 R}\gamma_\mu d_{3 R}\big{]}\big{[}\Bar{L}_3\gamma^\mu  L_3\big{]}\,,\label{eq:Old}
\end{align}
where $Q$ and $L$ denote the quark and lepton $SU(2)_L$ doublet, respectively, while $d$ stands for the down quark $SU(2)_L$ singlet\footnote{Within the SMEFT, there are also the $u$ and $e$ singlets. However, only down-type quarks for right-handed quark currents and left-handed lepton currents are involved in this study.}. We consider only the third generation in the lepton bilinears. Thus, we can simplify the notation for the Wilson coefficients by defining 
\begin{align}\label{eq:4f WCs}
\big{[}C_{\ell q}^{(1)}\big{]}_{2333} &\equiv C^{(1)}_{23}\,, \\[0.4em]
\big{[}C_{\ell q}^{(3)}\big{]}_{2333} &\equiv C^{(3)}_{23}\,,\\[0.4em]
\big{[}C_{\ell q}^{(3)}\big{]}_{3333} &\equiv C^{(3)}_{33}\,,\\[0.4em]
\big{[}C_{\ell d}\big{]}_{2333} &\equiv C^{\ell d}_{23}\,.
\end{align}

The decomposition in $SU(2)_L$ components is given by
\begin{equation} \label{eq: SMEFT lagrangian}
\begin{split}
    \mathcal{L}_{\text{eff}}^{\text{NP}}&= 2\left(V_{cs}C_{23}^{(3)}+V_{cb}C_{33}^{(3)}\right)(\bar{c}_L\gamma_\mu b_L)(\bar{\tau}_L\gamma^\mu\nu_{L\tau})\\
    & +\left(C_{23}^{(1)}+C_{23}^{(3)}\right)(\bar{s}_L\gamma_\mu b_L)(\bar{\tau}_L\gamma^\mu\tau_L)\\
    & +\left(C_{23}^{(1)}-C_{23}^{(3)}\right)(\bar{s}_L\gamma_\mu b_L)(\bar{\nu}_{L\tau}\gamma^\mu\nu_{L\tau})\\
    &+C_{23}^{\ell d}\,\big[(\bar{s}_R\gamma_\mu b_R)(\bar{\nu}_{L\tau}\gamma^\mu\nu_{L\tau})+(\bar{s}_R\gamma_\mu b_R)(\bar{\tau}_L\gamma^\mu\tau_L)\big]\,+{\rm h.c.\,}.
\end{split}
\end{equation}
Note that here we are working in the down-aligned basis, such that CKM elements appear in the charged-current operators. 

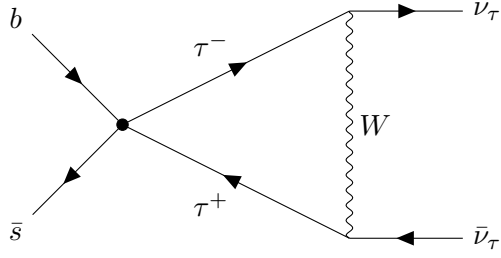
\begin{figure}[t] 
    \centering
    \begin{tikzpicture}
        \begin{feynman}
            \vertex [dot] (origin) {}; 
            \vertex [above left=2cm of origin] (b_in) {$b$};
            \vertex [below left=2cm of origin] (s_in) {$\bar{s}$};
            \vertex [right=3cm of origin, yshift=1.5cm] (v_top);
            \vertex [right=3cm of origin, yshift=-1.5cm] (v_bot);
            \vertex [right=1.5cm of v_top] (nu_out) {$\nu_\tau$};
            \vertex [right=1.5cm of v_bot] (nubar_out) {$\bar{\nu}_\tau$};
            \diagram* {
                (b_in) -- [fermion] (origin),
                (s_in) -- [anti fermion] (origin),
                (origin) -- [fermion, edge label=$\tau^-$] (v_top),    
                (origin) -- [anti fermion, edge label'=$\tau^+$] (v_bot),
               (v_top) -- [boson, edge label=$W$] (v_bot),
                (v_top) -- [fermion] (nu_out),
               (nubar_out) -- [fermion] (v_bot),
            };
        \end{feynman}
    \end{tikzpicture}
    \caption{Feynman diagram depicting the one-loop contribution generated by a mixing of ${\cal O}_9^\tau$ into ${\cal O}_{L,R}^{\nu_\tau\nu_\tau}$ generating an effect in $b\to s\nu\bar{\nu}$ processes even for $C^{(1)}_{23}=C^{(3)}_{23}$.}
    \label{fig: Feynman diagram W vertex}
\end{figure}

There is a $W$ vertex correction, shown in Fig.~\ref{fig: Feynman diagram W vertex}, generating a $\bar s b \bar \nu_\tau \nu_\tau$ operator out of an $\bar s b \bar \tau \tau$ one, linking both processes. 
The resulting extra contribution is
\begin{equation}
    C_{L}^{\nu_\tau\,\text{W}}=\frac{3}{4\sqrt{2}G_F\sin^2\theta_W}\frac{1}{V_{tb}V_{ts}^*}\log\Big(\frac{m_W^2}{\Lambda^2} \Big)\left(C_{23}^{(1)}+C_{23}^{(3)}\right)\,.
\end{equation}

In the following, we will consider different setups for these Wilson coefficients, which are now treated as corrections to the SM contributions in such a way that
\begin{align}
\label{C9 and C10}
C_{9(10)}^{\tau}&=C_{9(10)}^{\rm SM}+(-)\frac{1}{2\mathcal{N}}\left(C_{23}^{(1)}+C_{23}^{(3)}\right)\,,
&
C_{9'(10')}^{\tau}&=+(-)\frac{1}{2\mathcal{N}}C_{23}^{\ell d}\,,\\
\label{CL and CR}
C^{\nu_\tau}_{L}&=C_{L}^{\rm SM}+C_{L}^{\nu_\tau\, \text{W}}+\frac{1}{2\mathcal{N}}\left(C_{23}^{(1)}-C_{23}^{(3)}\right)\,,
&
C^{\nu_\tau}_{R}&=\frac{1}{2\mathcal{N}}C_{23}^{\ell d}\,,\\
\label{epsilonL}
\epsilon_L&=-\frac{1}{\sqrt{2}G_F}\left(\frac{V_{cs}}{V_{cb}}C_{23}^{(3)}+C_{33}^{(3)}\right)\,,
\end{align}
where 
\begin{equation}
    \mathcal{N}=\frac{4G_F}{\sqrt{2}}V_{tb}V_{ts}^*\frac{\alpha}{4\pi}\,,
\end{equation}
is the usual WET prefactor. Note that $C_{9'(10')}^{\rm SM}=C_{R}^{\rm SM}=0$ and $C_{L}^{\nu_\tau\, \text{W}} \propto\, C_{23}^{(1)}+C_{23}^{(3)}$.

Comparing the neutral (Eqs.~\eqref{C9 and C10} and ~\eqref{CL and CR}) and charged currents (Eq.~\eqref{epsilonL}), one can see the different scaling explicitly: In the SM, $b\to s$ transitions are generated at the loop level, whereas $b\to c$ transitions occur already at tree level. Consequently, NP contributions to FCNCs have an extra factor of $1/\alpha$, leading to corrections to the Wilson coefficients, which are two orders of magnitude larger than the relative effect in the charged currents. This factor is the clear signature of the distinct nature of FCCC  and FCNC, and the reason why small tensions w.r.t.~the SM in the former imply huge enhancements in the latter. 

\subsection{Phenomenological analysis}

We now turn to the SMEFT phenomenological analysis, focusing on the correlations among $b\to s\tau^+\tau^-$ observables, $R(D^{(*)})$, and $B\to K^{(*)}\nu\bar{\nu}$. In Ref.~\cite{Capdevila:2017iqn}, the analysis was performed under the assumptions
\begin{equation} \label{eq:basic}
    C_{23}^{(1)}=C_{23}^{(3)}
    \qquad {\rm and}\qquad
    C^{\ell d}_{23}=0\,,
\end{equation}
which imply the absence of tree-level contributions to $b\to s\nu\bar{\nu}$ transitions. We first revisit and update those results using the latest experimental and theoretical inputs and then extend the analysis beyond these assumptions and allow for a more general flavour structure. Although this introduces additional free parameters, the constraints from $B\to K^{(*)}\nu\bar{\nu}$ can be used to eliminate them, enabling us to derive predictions for $b\to s\tau^+\tau^-$ observables in terms of measured quantities.

In the following subsections, we will consider four different scenarios for the coefficients in Eq.~\eqref{eq: SMEFT lagrangian}.

\subsubsection{Scenario 1: $C_{23}^{(1)}=C_{23}^{(3)}$ and $|C_{23}^{(3)}V_{cs}|\gg |C_{33}^{(3)}V_{cb}|$}
\begin{figure}[t!]
    \centering
    \includegraphics[width=0.9\textwidth]{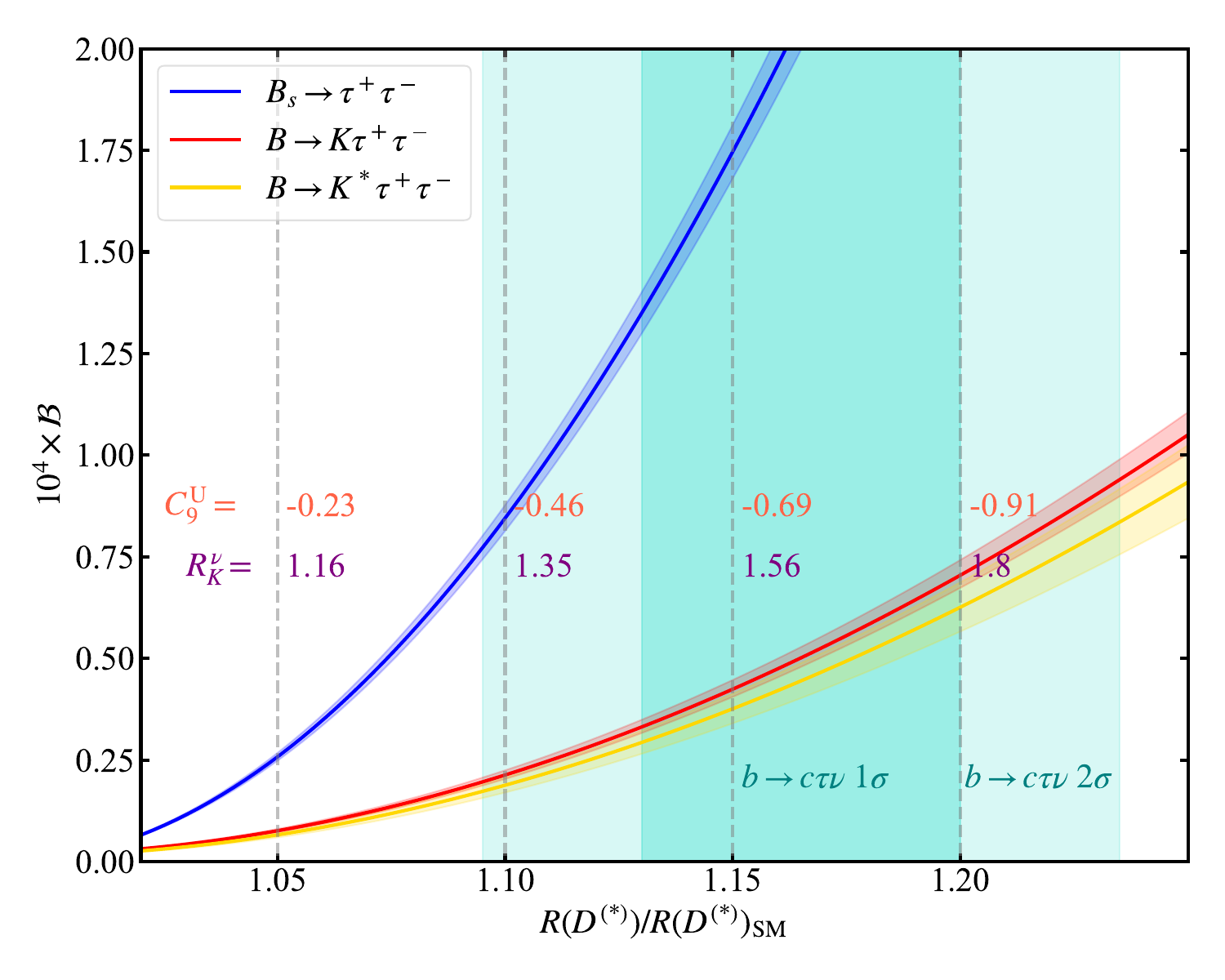}
    \caption{$b\to s\tau^+\tau^-$ processes as a function of $R(D^{(*)})/R(D^{(*)})_{\rm SM}$ in the scenario with $C_{23}^{(1)}=C_{23}^{(3)}$ and $|C_{23}^{(3)}V_{cs}|\gg |C_{33}^{(3)}V_{cb}|$. The contour lines denote the value of $C_9^{\rm U}$ and $
    R_{K}^\nu=
    R_{K^{*}}^\nu$ induced via the off-shell photon penguin and the $W$ vertex correction, respectively, for a NP scale $\Lambda=5$\,TeV. The dark (light) cyan region indicates the preferred $1\sigma$ ($2\sigma$) region from $b\to c\tau\nu$ measurements~\cite{Iguro:2024hyk}.}
    \label{fig:BRs vs RD}
\end{figure}

As we are working in the down-basis, $b\to c\tau\nu$ transitions are not only generated by $C^{(3)}_{23}$ but also by $C^{(3)}_{33}$ via CKM rotations in the process of EW symmetry breaking (see Eq.~(\ref{eq: SMEFT lagrangian})). However, for $|C_{23}^{(3)}V_{cs}|\gg |C_{33}^{(3)}V_{cb}|$ the effect of these rotations can be neglected. If we define
\begin{equation}
    C_{23}\equiv C_{23}^{(1)}=C_{23}^{(3)}\,,
\end{equation}
to parametrise NP effects in the absence of a tree-level effect in $b\to s\nu\bar{\nu}$ transitions, we can express $C_{23}$ in terms of $R(D^{(*)})$ to obtain\footnote{Here, we choose the solution in which the NP effect does not overcompensate the SM, since the other solution would require large couplings and thus would conflict with the limits from LHC searches for high-$p_T$ single tau leptons~\cite{Greljo:2018tzh}. By overcompensation, we mean that the NP contribution is at least twice the absolute value of the SM Wilson coefficient and interferes destructively.}~\cite{Capdevila:2017iqn}
\begin{equation}\label{eq:C910tautau_old}
C_{9(10)}^{\tau}=C_{9(10)}^{\text{SM}}-(+)\Delta_{R(D^{(*)})}\,,
\end{equation}
with
\begin{equation}
\label{DeltaRD}
    \Delta_{R(D^{(*)})}\equiv\frac{2\pi}{\alpha}\frac{V_{cb}}{V_{cs}V_{tb}V_{ts}^*}\left(\sqrt{\frac{R(D^{(*)})}{R(D^{(*)})_{\rm SM}}}-1\right)\,.
\end{equation}
Note that $V_{cb}/(V_{cs}V_{tb}V_{ts}^*)=-1+\mathcal{O}(\lambda^2)$ expanding on the $\lambda$ Wolfenstein parameter. 
Furthermore, in terms of $R(D^{(*)})/R(D^{(*)})_{\rm SM}$ the effect in the neutrino mode (the $W$ vertex correction) and $C_9^{\rm U}$ (off-shell photon penguin), discussed in the last section, are given by
\begin{equation} \label{eq: Rknu scenario 1}
    {R_{K}^\nu=R_{K^{*}}^\nu}  
    =\frac{2}{3}+\frac{1}{3}\left\lvert1+\frac{1}{C_L^{\text{SM}}}\frac{3}{2\sin^2\theta_W}\frac{V_{cb}}{V_{cs}V_{tb}V_{ts}^*}\log\bigg(\frac{m_W^2}{\Lambda^2}\bigg)\Bigg(1-\sqrt{\frac{R(D^{(*)})}{R(D^{(*)})_{\rm SM}}}\Bigg)\right\rvert^2\\,
\end{equation}
and
\begin{equation} \label{eq: C9U scenario 1}
    C_9^{\rm U}=\frac{2}{3}\frac{V_{cb}}{V_{cs}V_{tb}V_{ts}^*}\log\left(\frac{\Lambda^2}{\mu_b^2}\right)\Bigg(\sqrt{\frac{R(D^{(*)})}{R(D^{(*)})_{\rm SM}}}-1\Bigg)\,,
\end{equation}
respectively.

This relation is illustrated in Fig.~\ref{fig:BRs vs RD} for the different $b\to s\tau^+\tau^-$ processes under consideration. As one can see, the branching ratios of semileptonic $B$ meson decays to tau leptons are predicted to be enhanced to the level of $10^{-5}$ within the region currently preferred by $b\to c\tau\nu$ data. Furthermore, we show the predictions for $C_9^{\rm U}$ generated via the off-shell tau loop, which can (partially) explain the anomalies in $b\to s\ell^+\ell^-$ observables, and the impact of the $W$ vertex correction generating a loop effect in $B\to K^{(*)}\nu\bar{\nu}$ via the diagram in Fig.~\ref{fig: Feynman diagram W vertex}. 

\subsubsection{Scenario 2: $C_{23}^{(1)}\neq C_{23}^{(3)}$ and $|C_{23}^{(3)}V_{cs}|\gg |C_{33}^{(3)}V_{cb}|$}

\begin{figure}[htbp]
    \centering
\includegraphics[width=0.8\textwidth]{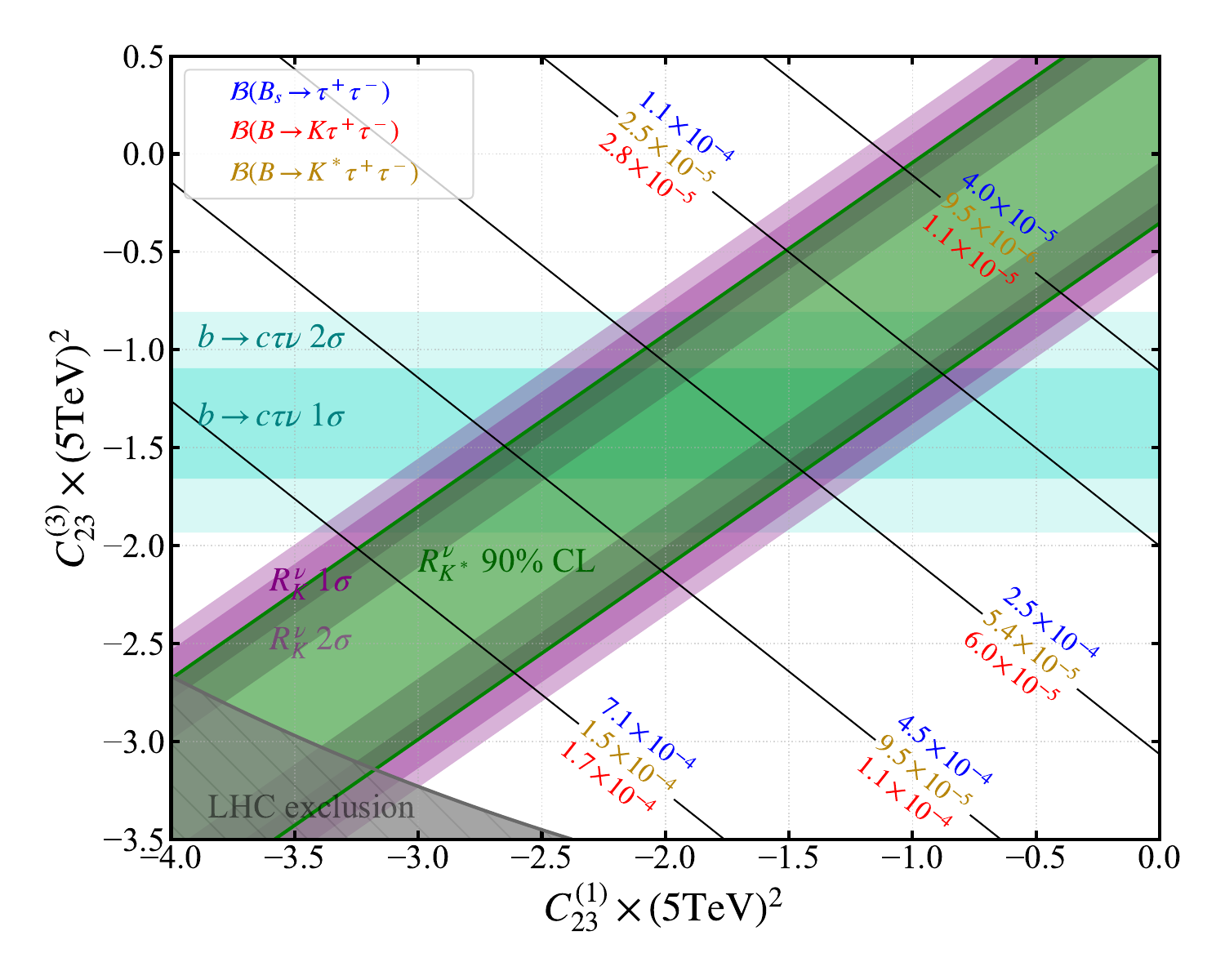}
    \caption{Preferred and excluded regions in the $C_{23}^{(1)}$--$C_{23}^{(3)}$ plane for a NP scale $\Lambda=5$\,TeV. The horizontal cyan band is preferred by $b\to c\tau\nu$ measurements, the purple strips represent regions preferred by $R_K^\nu$~\cite{Belle-II:2023esi}, while the green region is allowed by $R_{K^*}^\nu$ at $90\%$ CL~\cite{Belle:2017oht}. The hatched region on the bottom-left is excluded by non-resonant mono-tau searches at the LHC. The black lines show constant values for $\mathcal{B}(B_s\to \tau^+\tau^-)$ (blue), $\mathcal{B}(B\to K\tau^+\tau^-)$ (red) and $\mathcal{B}(B\to K^{*}\tau^+\tau^-)$ (dark yellow). Note that there are two diagonal regions preferred by $R_K^\nu$ where the upper band corresponds to an overcompensation of the SM contribution to tau neutrinos by NP. For the same value of $C_{23}^{(3)}$ (i.e.~$R(D^{(*)})$), the solution with overcompensation leads to a larger effect in $b\to s\tau^+\tau^-$ processes by a factor of $\approx 2$.}
    \label{fig:C1-Cld plane.}
\end{figure}
\begin{figure}[htbp]
    \centering
    \begin{subfigure}[b]{0.48\textwidth}
        \centering
        \includegraphics[width=\textwidth]{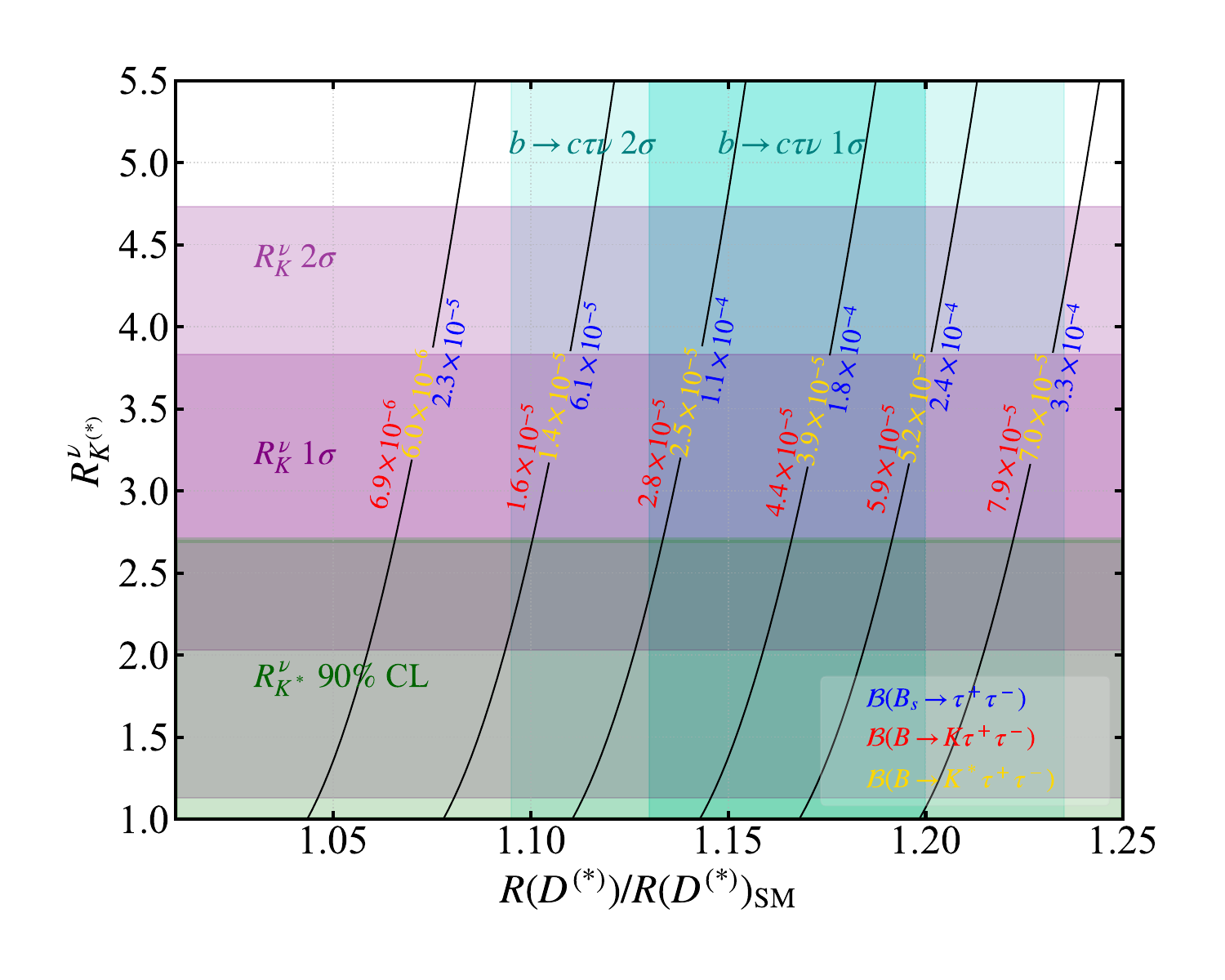}
        \caption{NP interferes constructively with the SM.}
    \end{subfigure}
    \hfill 
    \begin{subfigure}[b]{0.48\textwidth}
        \centering
        \includegraphics[width=\textwidth]{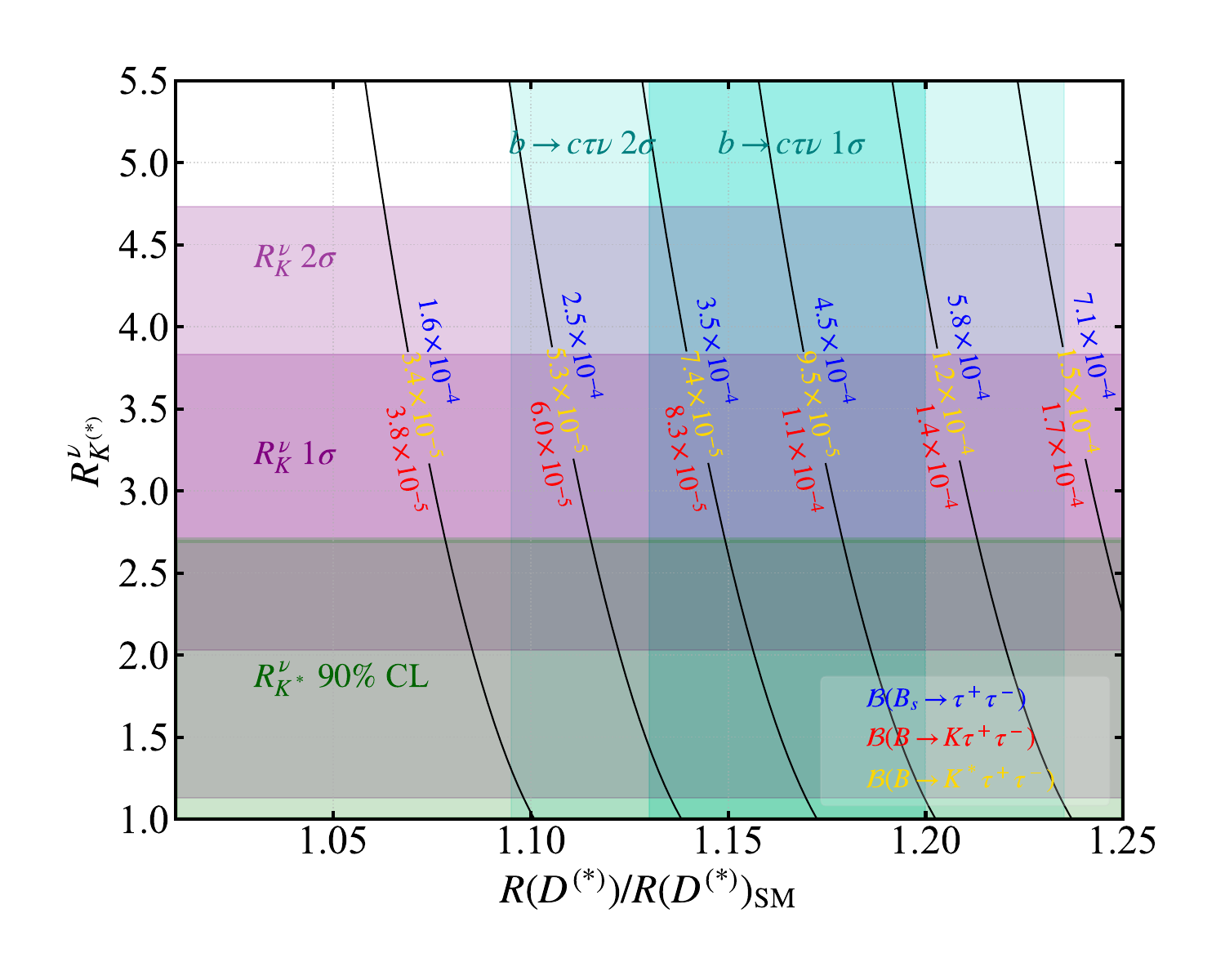}
        \caption{NP interferes destructively with the SM.}
    \end{subfigure}
    \caption{Comparison between the two different solutions for the theory predictions of $\mathcal{B}(B_s\to \tau^+\tau^-)$ (blue) and $\mathcal{B}(B\to K^{(*)}\tau^+\tau^-)$ (red (yellow)), in the $R_{K^{(*)}}^\nu$--$R(D^{(*)})/R(D^{(*)})_{\rm SM}$ plane: the one where NP interferes constructively with the SM (left) in $B\to K^{(*)}\nu\bar{\nu}$ and the one where NP overcompensates (right) the SM contribution. Vertical light blue bands are preferred by $b\to c\tau\nu$ measurements, horizontal purple bands indicate the region preferred by $R_K^\nu$, while the green region is consistent with $R_{K^*}^\nu$ at $90\%$ CL.}\label{fig:comparison_solutions_RKnu_RD_plane}
    \end{figure}
\begin{figure}[htbp]
    \centering
    \includegraphics[width=0.8\textwidth]{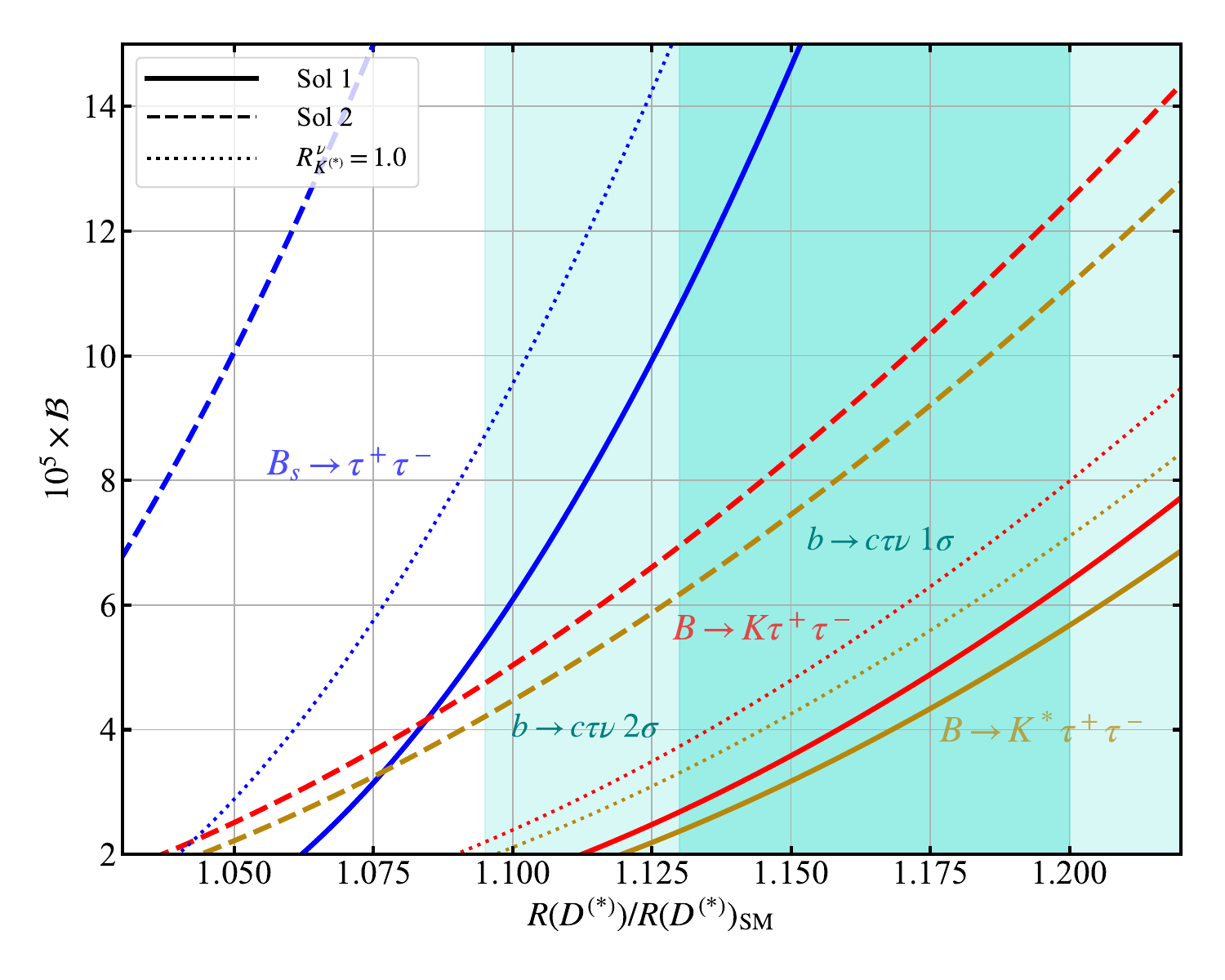}
    \caption{$\mathcal{B}(B_s\to \tau^+\tau^-)$ (blue), and $\mathcal{B}(B\to K^{(*)}\tau^+\tau^-)$ (red(yellow)) as a function of $R(D^{(*)})/R(D^{(*)})_{\rm SM}$ for $R_{K^{(*)}}^\nu=2.9$ (current central value of $R_K^\nu$ averaging over all measurements up to date). For solution~1 (solid), NP interferes constructively with the SM in $B\to K^{(*)}\nu\bar{\nu}$, while for solution~2 (dashed) we have destructive interference. The cyan region is preferred by the $b\to c\tau\nu$ measurements. The dotted lines represent the scenario with no NP in $B\to K^{(*)}\nu\bar \nu$, which corresponds to $R_{K^{(*)}}^\nu=1$ for solution 1.}
    \label{fig:comparison_solutions_vs_RD}
\end{figure}
    
We now allow for $C_{23}^{(1)}\neq C_{23}^{(3)}$, such that $b\to s\nu\bar{\nu}$ transitions can already be generated at tree level, but still consider the case of left-handed currents only such that $R_K^\nu=R_{K^*}^\nu$. Even though we have one more free parameter now, we can still express $C_{9(10)}^{\tau\,{\rm NP}}$ in terms of both $b\to c\tau\nu$ and $b\to s\nu\bar{\nu}$ observables: $C^{(3)}_{23}$ can be eliminated by $R(D^{(*)})$ and $C^{(1)}_{23}-C^{(3)}_{23}$ can be expressed in terms of  $B\to K^{(*)}\nu\bar{\nu}$. However, note that here there is more than one solution. Since the possible size of an effect in $b\to s\nu\bar{\nu}$ is weakly constrained, it is possible that NP interferes destructively with the SM contribution to $C^{\nu_\tau}_{L}$ (see Eq.~\eqref{CL and CR}), and overcompensates it, resulting in the following two solutions\footnote{
The existence of these two solutions stems from the fact that two different  values of $C_{23}^{(1)}$ and $C_{23}^{(3)}$
yield the same value of $R_{K^{(*)}}^\nu$. In contrast, these two values give rise to substantially different contributions to 
$C_{9(10)}^{\tau}$ (see Eq.~\eqref{C9 and C10}) and consequently lead to distinct predictions for the $b\to s \tau^+\tau^-$
 observables.}
(for $R_{K^{(*)}}^\nu>1$):
\begin{itemize}
    \item {\bf{Solution 1}}: NP interferes constructively with the SM in $B\to K^{(*)}\nu\bar{\nu}$, such that
    \begin{equation}\label{eq: C910tautau C1!=C3 only LHC, sol+}
C_{9(10)}^{\tau}=C_{9(10)}^{\text{SM}}-(+)\left(\Tilde{\Delta}_{R(D^{(*)})}-\Tilde{\Delta}^+_{R_{K^{(*)}}^\nu}\right)\\ ,
    \end{equation}
    \item {\bf{Solution 2}}: NP interferes destructively with the SM and overcompensates its effect in $B\to K^{(*)}\nu\bar{\nu}$, such that
    \begin{equation}\label{eq: C910tautau C1!=C3 only LHC, sol-}
C_{9(10)}^{\tau}=C_{9(10)}^{\text{SM}}-(+)\left(\Tilde{\Delta}_{R(D^{(*)})}-\Tilde{\Delta}^-_{R_{K^{(*)}}^\nu}\right)\\ ,
    \end{equation}
\end{itemize}
where the $W$ vertex correction appears explicitly through the rescaling prescription
\begin{equation}
\label{W vertex effect}
\Tilde{X}=\dfrac{1}{1+\kappa}X\,,
\qquad 
\kappa=\dfrac{\alpha}{\pi}\dfrac{3}{4\sin^2\theta_W}\log\Big(\dfrac{m_W^2}{\Lambda^2}\Big)\,,
\end{equation}
for $X=\Delta_{R(D^{(*)})},\,\Delta^\pm_{R_{K^{(*)}}^\nu}$.
The two solutions mentioned above are given by
\begin{equation}
\label{DeltaRKnu}    \Delta^\pm_{R_{K^{(*)}}^\nu}=C_L^{\text{SM}}\left(\pm\sqrt{3R_{K^{(*)}}^\nu-2}-1\right).
\end{equation}

In Fig.~\ref{fig:C1-Cld plane.}, we show the predictions for $b\to s\tau^+\tau^-$ channels in the $C_{23}^{(1)}-C_{23}^{(3)}$ plane. The purple strips in that figure correspond to the regions preferred by $b\to s\nu\bar{\nu}$ measurements\footnote{$R_K^\nu$ and $R_{K^*}^\nu$ have been calculated using the SM predictions from \cite{Becirevic:2023aov} and \cite{Buras:2014fpa}, respectively. The former value does not include the tree-level contribution~\cite{Kamenik:2009kc} which is subtracted in the experimental analysis~\cite{Belle-II:2023esi}.}. LHC searches for mono-tau signatures~\cite{ATLAS:2019lsy,ATLAS:2024tzc,Greljo:2018tzh} exclude the hatched region on the bottom-left\footnote{We used HighPT~\cite{Allwicher:2022mcg} throughout this work to calculate LHC exclusion regions in all figures involving SMEFT Wilson coefficients.}. The solution where NP overcompensates the SM in $B\to K^{(*)}\nu\bar{\nu}$ corresponds to the upper purple band, which results in a more significant enhancement of the $b\to s\tau^+\tau^-$ modes than the case of constructive interference (lower purple band). Furthermore, in Fig.~\ref{fig:comparison_solutions_RKnu_RD_plane} we show the predictions for the various $b\to s\tau^+\tau^-$ processes in the $R(D^{(*)})-R_{K^{(*)}}^\nu$ plane for the two solutions, i.e.~constructive interference (left) and destructive interference (right). Also here, note that for the same value of $R(D^{(*)})$ and $R_{K^{(*)}}^\nu$, the destructive case predicts larger rates for the $b\to s\tau^+\tau^-$ processes. Finally, in Fig.~\ref{fig:comparison_solutions_vs_RD} we show $\mathcal{B}(B_s\to \tau^+\tau^-)$ and $\mathcal{B}(B\to K^{(*)}\tau^+\tau^-)$ for $R_{K^{(*)}}^\nu=2.9$  as a function of $R(D^{(*)})$ for both solution 1 and solution 2. One can see that the presence of NP in $R_{K^{(*)}}^\nu$ implies an enhancement or reduction (compared to the case where there is no NP in $b\to s\nu\bar{\nu}$ and $R_{K^{(*)}}^\nu=1$, shown in dotted lines) of the $b\to s\tau^+\tau^-$ modes depending on whether there is a destructive or constructive interference with the SM, respectively.

\subsubsection{Scenario 3: $C_{23}^{(1)}\neq C_{23}^{(3)}$, $|C_{23}^{(3)}V_{cs}|\gg |C_{33}^{(3)}V_{cb}|$, and $C_{23}^{\ell d}\neq 0$}

\begin{figure}[htbp]
    \centering
\includegraphics[width=0.8\textwidth]{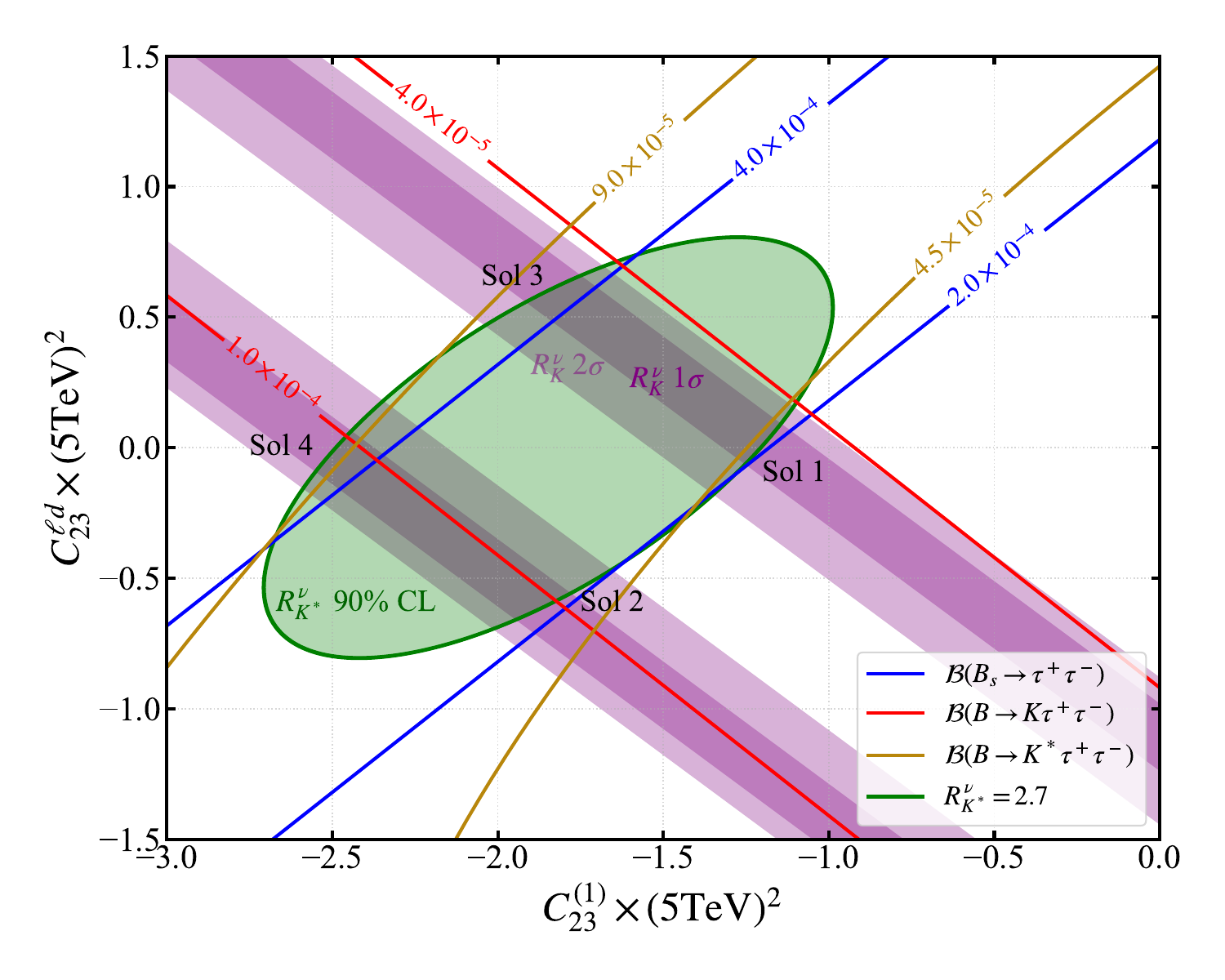}
    \caption{Preferred regions in the $C_{23}^{(1)}$--$C_{23}^{\ell d}$ plane for $C_{23}^{(3)}=-1.4/(5\,{\rm TeV})^2$ (corresponding to $R(D^{(*)})/R(D^{(*)})_{\rm SM}=1.17$). The purple strips are preferred by $R_K^\nu$, while the green region agrees with $R_{K^*}^\nu$ at $90\%$~CL. The solid contour lines show $\mathcal{B}(B_s\to \tau^+\tau^-)$ and $\mathcal{B}(B\to K^{(*)}\tau^+\tau^-)$. For a given value of $R_{K^*}^\nu$ ($R_{K^*}^\nu =2.7$ in the figure, accounting for the experimental upper bound), there are four possible combinations of WCs fulfilling $R_K^\nu$ experimental constraints, indicated as solutions 1 to 4 (see Table~\ref{tab:solutions}). Note that the bounds from non-resonant mono-tau searches at the LHC lie outside the plotted range. }
    \label{fig:C1-Cld plane RHC.}
\end{figure}
\begin{figure}[htbp]
    \centering
    \begin{subfigure}[b]{0.48\textwidth}
        \centering
        \includegraphics[width=\textwidth]{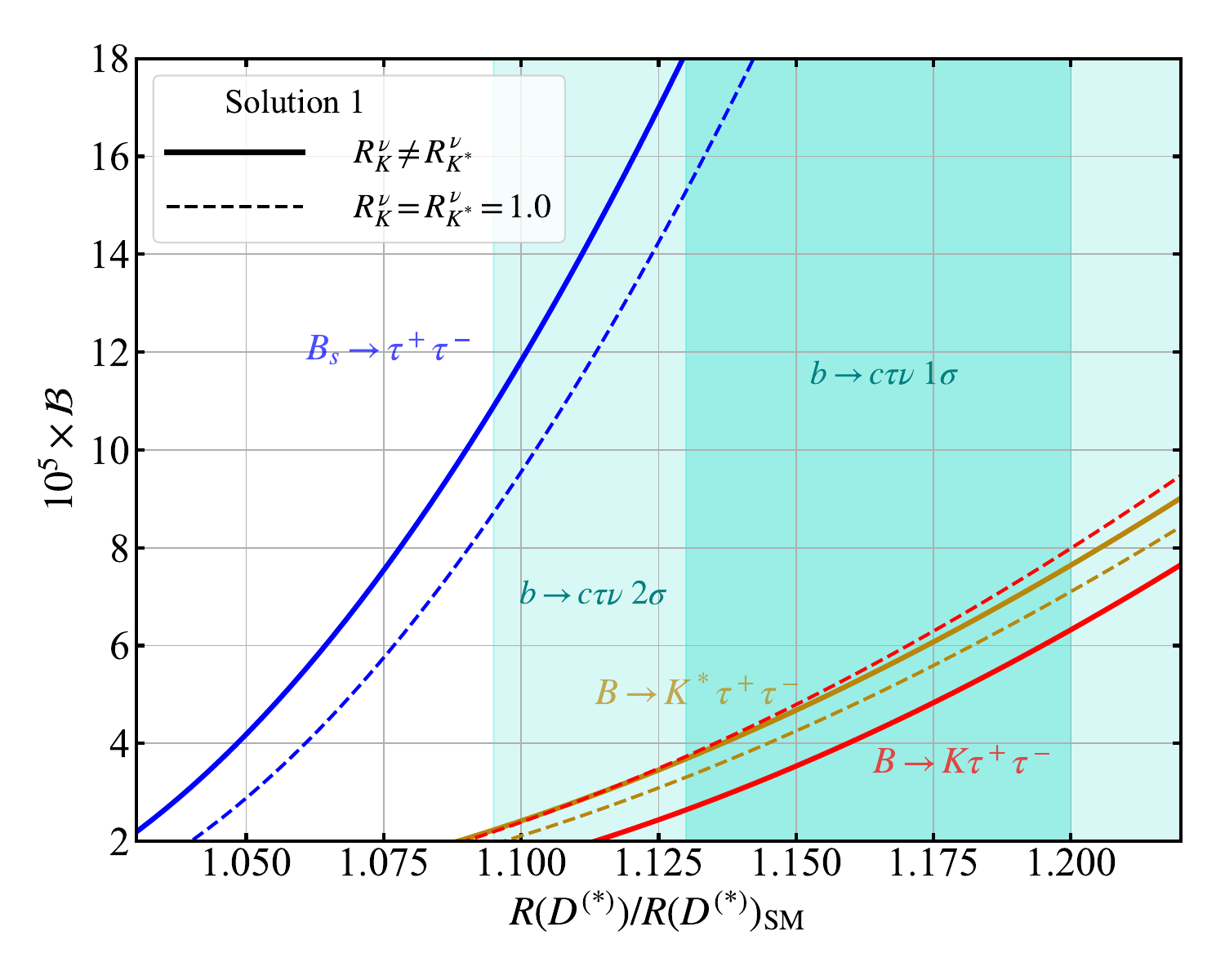}
        \caption{Solution 1}
        \label{fig:sol_1}
    \end{subfigure}
    \hfill 
    \begin{subfigure}[b]{0.48\textwidth}
        \centering
        \includegraphics[width=\textwidth]{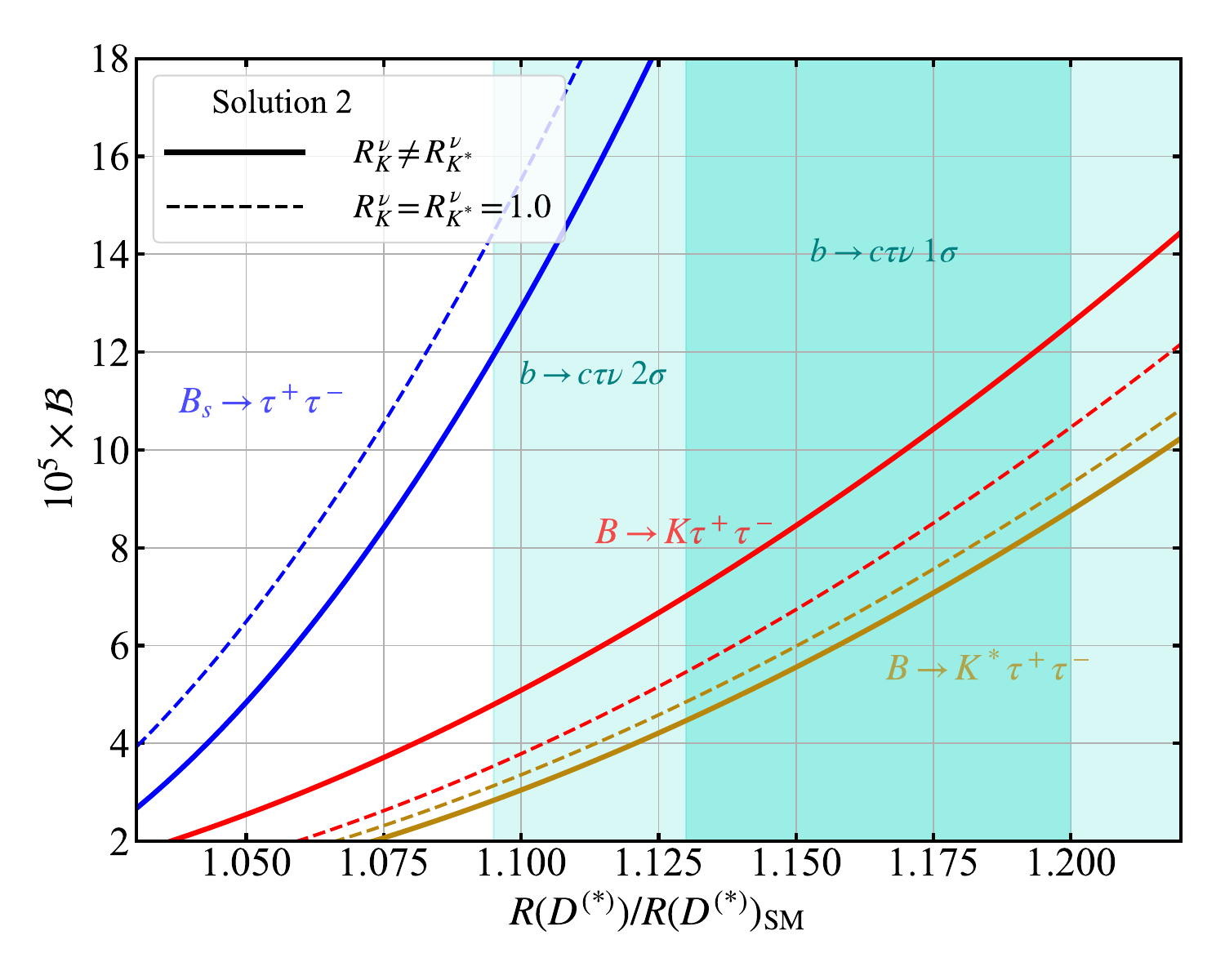}
        \caption{Solution 2}
        \label{fig:sol_2}
    \end{subfigure}
    \begin{subfigure}[b]{0.48\textwidth}
        \centering
        \includegraphics[width=\textwidth]{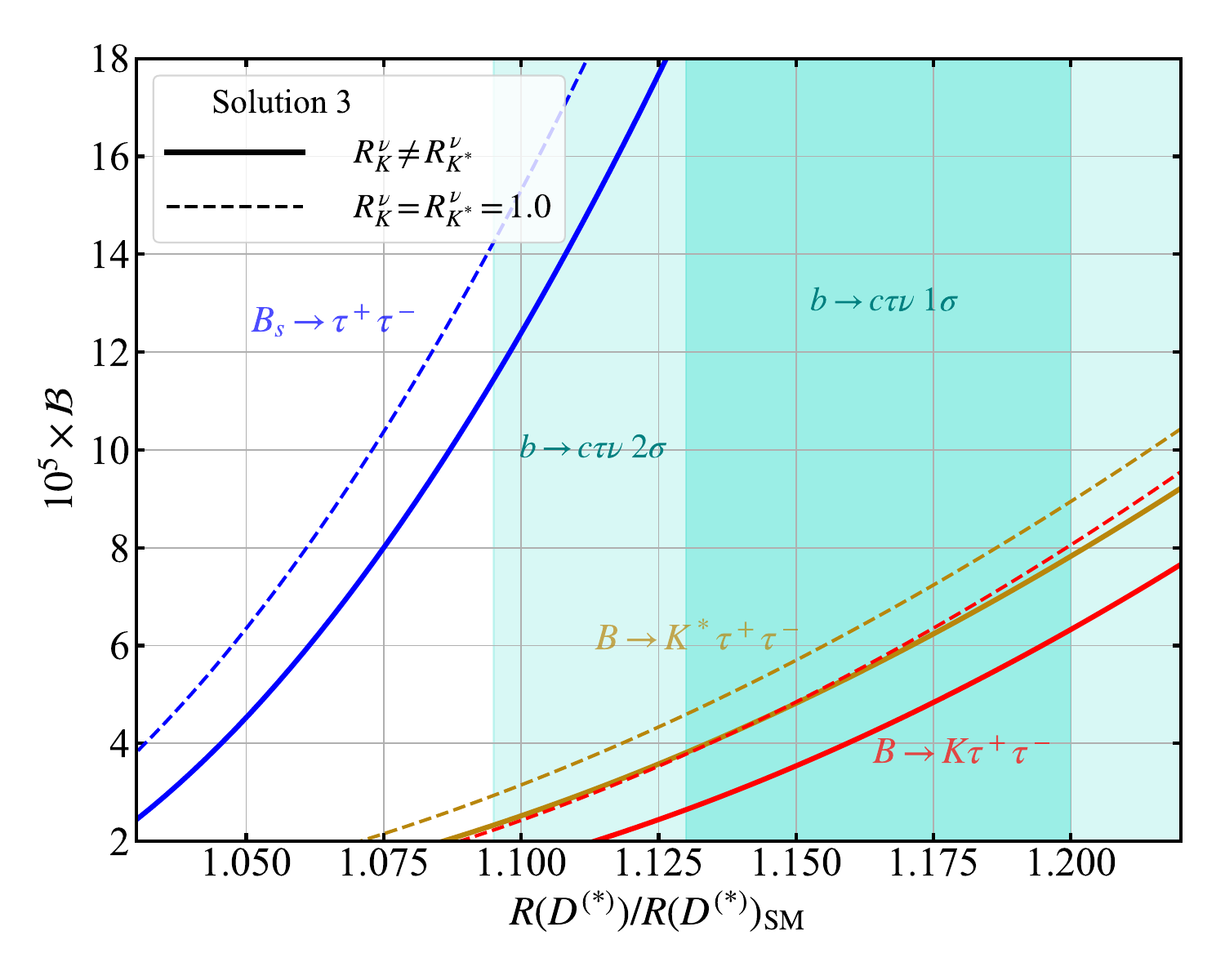}
        \caption{Solution 3}
        \label{fig:sol_3}
    \end{subfigure}
    \begin{subfigure}[b]{0.48\textwidth}
        \centering
        \includegraphics[width=\textwidth]{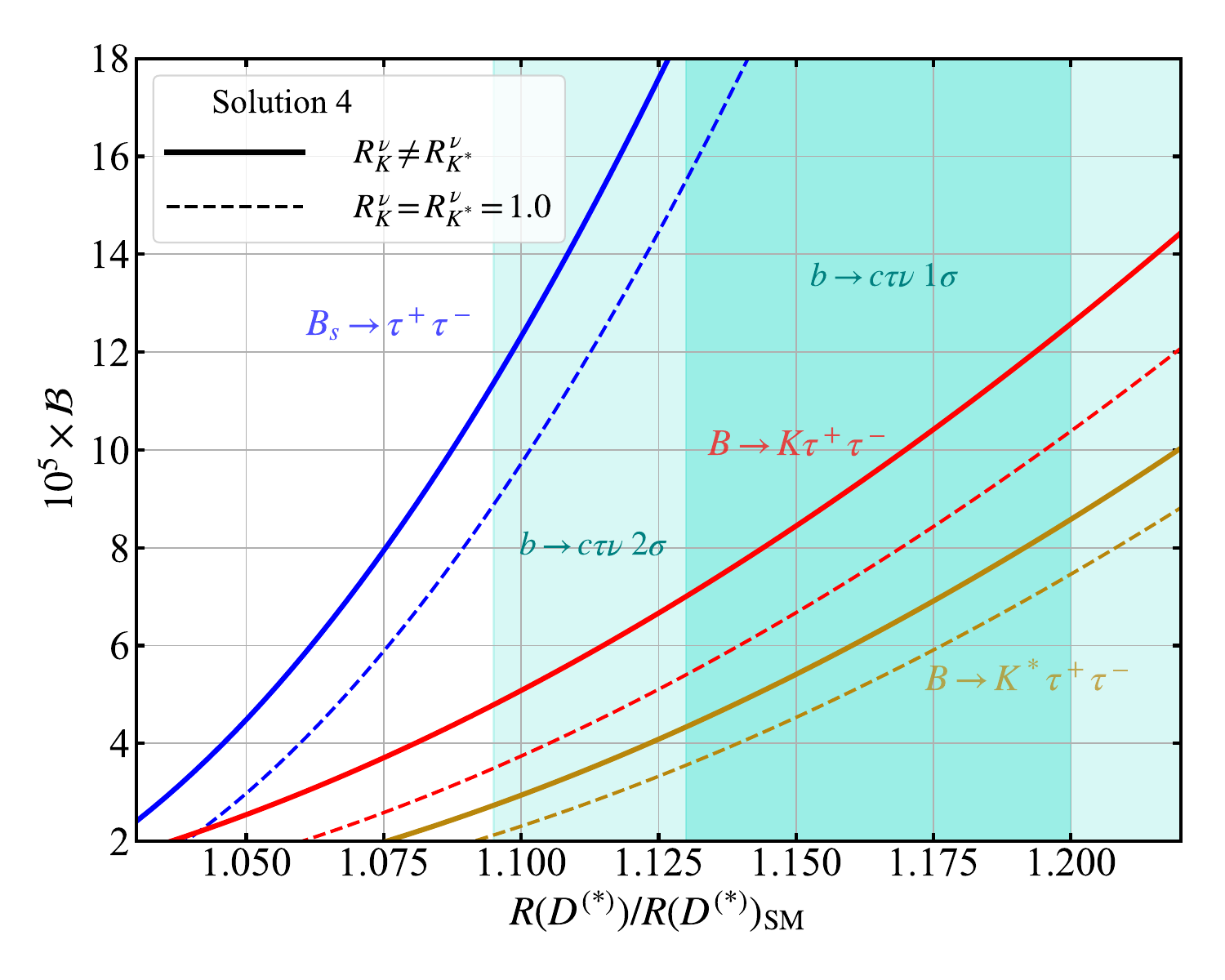}
        \caption{Solution 4}
        \label{fig:sol_4}
    \end{subfigure}
    \caption{Comparison of $\mathcal{B}(B_s\to \tau^+\tau^-)$ (blue), $\mathcal{B}(B\to K^{}\tau^+\tau^-)$ (red) and $\mathcal{B}(B\to K^{*}\tau^+\tau^-)$ (yellow), as a function of $R(D^{(*)})/R(D^{(*)})_{\rm SM}$ for the four different solutions (see Table~\ref{tab:solutions}) in the scenario including right-handed $\bar s b$ currents. The light-blue region is preferred by the $b\to c\tau\nu$ measurements. Solid lines represent the case $R_K^\nu=2.9$ (current measured central value) and $R_{K^*}^\nu=1.1$ (minimum theoretically allowed value according to Eqs.~\eqref{eq: Delta_pm} and~\eqref{eq: Delta_RHC}). Dashed lines depict the case where $R_{K^{(*)}}^\nu=1$. Note that this only corresponds to the case of no NP in $b\to s\nu\bar \nu$ processes in solution 1.}
    \label{fig:comparison_solutions_scn3}
\end{figure}

In the presence of left-handed currents only, $R_K^\nu=R_{K^*}^\nu$. However, in light of current data, this might not necessarily be the case since only $B\to K\nu\bar\nu$ shows an excess, while for $B\to K^*\nu\bar\nu$ the upper limit is slightly stronger than expected. Thus, we include right-handed quark currents (RHCs) in the analysis. This means $C_{23}^{\ell d}\neq 0$, which leads to a NP contribution to $C^{\nu_\tau}_{R}$ (see Eq.~\eqref{CL and CR}). In this setup, one can express the $b\to s\tau^+\tau^-$ Wilson coefficients, and thus the corresponding rates, in terms of $R(D^{(*)})$, $R_{K}^{\nu}$ and $R_{K^*}^{\nu}$. Since the presence of RHC implies differences between the two latter observables, we now have a system of three equations that needs to be solved. Note that there are four solutions due to the possibility of NP interfering destructively with the SM and overcompensating it in $R_K^\nu$ and/or $R_{K^*}^\nu$:

\bigskip
    
 \noindent $\bullet$   {\bf Solution 1}: 
    \begin{equation}  \biggr\{C_{9(10)}^{\tau},C_{9'(10')}^{\tau}\biggr\}=\biggr\{C_{9(10)}^{\text{SM}}-(+)\left(\Tilde{\Delta}_{R(D^{(*)})}-\Tilde{\Delta}^+_{R_{K^{(*)}}^\nu}\right), +(-)\delta_{R_{K^{(*)}}^\nu}\biggr\}\,.
    \end{equation}

 \noindent $\bullet$   {\bf Solution 2}:
    \begin{equation}
\biggr\{C_{9(10)}^{\tau},C_{9'(10')}^{\tau}\biggr\}=\biggr\{C_{9(10)}^{\text{SM}}-(+)\left(\Tilde{\Delta}_{R(D^{(*)})}-\Tilde{\Delta}^-_{R_{K^{(*)}}^\nu}\right), -(+)\delta_{R_{K^{(*)}}^\nu}\biggr\}\,.
    \end{equation}

\noindent $\bullet$    {\bf Solution 3}:
    \begin{equation}
\biggr\{C_{9(10)}^{\tau},C_{9'(10')}^{\tau}\biggr\}=\biggr\{C_{9(10)}^{\text{SM}}-(+)\left(\Tilde{\Delta}_{R(D^{(*)})}-\Tilde{\delta}_{R_{K^{(*)}}^\nu}+\Tilde{C}_L^{\rm SM}\right), +(-)\left(\Delta^+_{R_{K^{(*)}}^\nu}+C_L^{\rm SM}\right)\biggr\}\,.
    \end{equation}

 \noindent $\bullet$   {\bf Solution 4}:
    \begin{equation}
\biggr\{C_{9(10)}^{\tau},C_{9'(10')}^{\tau}\biggr\}=\biggr\{C_{9(10)}^{\text{SM}}-(+)\left(\Tilde{\Delta}_{R(D^{(*)})}+\Tilde{\delta}_{R_{K^{(*)}}^\nu}+\Tilde{C}_L^{\rm SM}\right), -(+)\left(\Delta^+_{R_{K^{(*)}}^\nu}+C_L^{\rm SM}\right)\biggr\}\,.
    \end{equation}
    
Here, the function $\Delta_{R(D^{(*)})}$ is the same as in the previous sections (see Eq.~\eqref{DeltaRD}), but the analogue of Eq.~\eqref{DeltaRKnu} becomes
\begin{equation}
\label{eq: Delta_pm}
\Delta^{\pm}_{R_{K^{(*)}}^\nu}= C_L^{\rm SM}\left(\pm\sqrt{\frac{3}{2} (\hat{R}_K -\Delta\hat{R}) \left[1 + \sqrt{1 -\frac{{\Delta\hat{R}}^2}{
(\hat{R}_K- \Delta\hat{R})^2}
}\right]}-1\right)\,.
\end{equation}
Additionally, a new term $\delta_{R_{K^{(*)}}^\nu}$ appears defined as
\begin{equation}
\label{eq: Delta_RHC}
\delta_{R_{K^{(*)}}^\nu}= C_L^{\rm SM}\sqrt{\frac{3}{2} (\hat{R}_K -\Delta\hat{R}) \left[1 - \sqrt{1 -\frac{{\Delta\hat{R}}^2}{
(\hat{R}_K- \Delta\hat{R})^2}
}\right]}\,,
\end{equation}
with
\begin{equation}
    \hat{R}_K=R_K^{\nu}-2/3\,,\qquad \Delta \hat{R}= 2 (R_K^{\nu}-R_{K^*}^{\nu})/\eta_{K^\ast}\,,
\end{equation}
where $\Delta\hat{R}$ encodes the effect of RHC (i.e. $C_{23}^{\ell d}\neq 0$).
Finally, $\Tilde{\Delta}_{R(D^{(*)})}$, $\Tilde{\Delta}^{\pm}_{R_{K^{(*)}}^\nu}$, $\Tilde{\delta}_{R_{K^{(*)}}^\nu}$, and $\Tilde{C}_L^{\rm SM}$ are defined following Eq.~\eqref{W vertex effect} to account for the W vertex correction.  Nevertheless, the numerical impact is only significant for the shift in $\Tilde{\Delta}_{R(D^{(*)})}$ relative to its uncorrected counterpart, $\Delta_{R(D^{(*)})}$.

\begin{table}[ht]
\centering
\vspace{0.5em}
\begin{tabular}{| c | >{\centering\arraybackslash}p{3.5cm} | >{\centering\arraybackslash}p{3.5cm} |}
\hline
 & Destructive interference with the SM in $B\to K\nu\bar{\nu}$ & Destructive interference with the SM in $B\to K^{*}\nu\bar{\nu}$ \\ \hline
Sol 1 & $\times$ & $\times$ \\ \hline
Sol 2 & $\checkmark$ & $\times$ \\ \hline
Sol 3 & $\times$ & $\checkmark$ \\ \hline
Sol 4 & $\checkmark$ & $\checkmark$ \\ \hline
\end{tabular}
\caption{Different solutions for $b\to s\tau^+\tau^-$ processes in terms of $R_{K^{(*)}}^{\nu}$ and $R(D^{(*)})$ as shown in Fig.~\ref{fig:comparison_solutions_scn3} for $R_{K^{(*)}}^\nu>1$. They are classified depending on whether NP interferes constructively or destructively with the SM in $B\to K\nu\bar{\nu}$ and $B\to K^{*}\nu\bar{\nu}$ modes. Note that we do not consider solutions where NP interferes destructively with the SM in $b\to c\tau\nu$, as this would require huge effects which are in general in conflict with direct LHC searches.
\label{tab:solutions}}
\end{table}

While only solution 1 can be continuously obtained from the SM limit if $R_K^\nu>1$ and $R_{K^*}^\nu>1$, the other solutions involve an overcompensation of the SM in tau neutrino mode in $R_K^\nu$ and/or $R_{K^*}^\nu$. The four different solutions are classified following this reasoning in Table~\ref{tab:solutions} and seen in Fig.~\ref{fig:C1-Cld plane RHC.} by the four overlapping regions of the currently preferred bands from $R_K^\nu$ and $R_{K^*}^\nu$.

The predictions for the $b\to s\tau^+\tau^-$ rates are shown in Fig.~\ref{fig:comparison_solutions_scn3} for $R_K^\nu=2.9$ and $R_{K^*}^\nu=1.1$ as a function of $R(D^{(*)})/R(D^{(*)})_{\rm SM}$. The latter value is the minimum theoretically allowed for the corresponding current measured central value of $R_K^\nu$, leading to the maximum possible effect in $b\to s\tau^+\tau^-$ channels. The effect is represented in the four solutions compared to the prediction for $R_{K^{(*)}}^\nu=1$\footnote{Note that in solution 1, which is the only solution with constructive interference between SM and NP in both $B\to K\nu\bar{\nu}$ and $B\to K^{*}\nu\bar{\nu}$, $R_{K^{(*)}}^\nu=1$ means the absence of NP in those observables. However, for the rest of the solutions this is not the case.}. The strongest enhancement is obtained with solution 2, while in solution 1 the enhancement of $b\to s\tau^+\tau^-$ processes is the weakest. 

The fact that four solutions exist is especially interesting from the experimental point of view. The theoretical phenomenon of having several solutions for the same  measured values of the observable happens in $R(D^{(*)})/R(D^{(*)})_{\rm SM}$ too, with two solutions appearing. But, as we mentioned previously in this article, in that case the second solution can be excluded by LHC searches for high-$p_T$ single tau leptons. Such experimental exclusions do not exist for $R_{K^{(*)}}^\nu$, meaning that $b\to s\tau^+\tau^-$ searches provide a handle on disentangling the overcompensation solutions from cases in which there is constructive interference with the SM in $R_K^\nu$ and $R_{K^*}^\nu$. Thus, they play an important role in probing NP in the neutrino modes.

\subsubsection{Scenario 4: $|C_{33}^{(3)}|\gg |C_{23}^{(3)}|$}

\begin{figure}[htbp]
    \centering
    \begin{subfigure}[b]
    {0.75\textwidth}
        \centering
        \includegraphics[width=\textwidth]{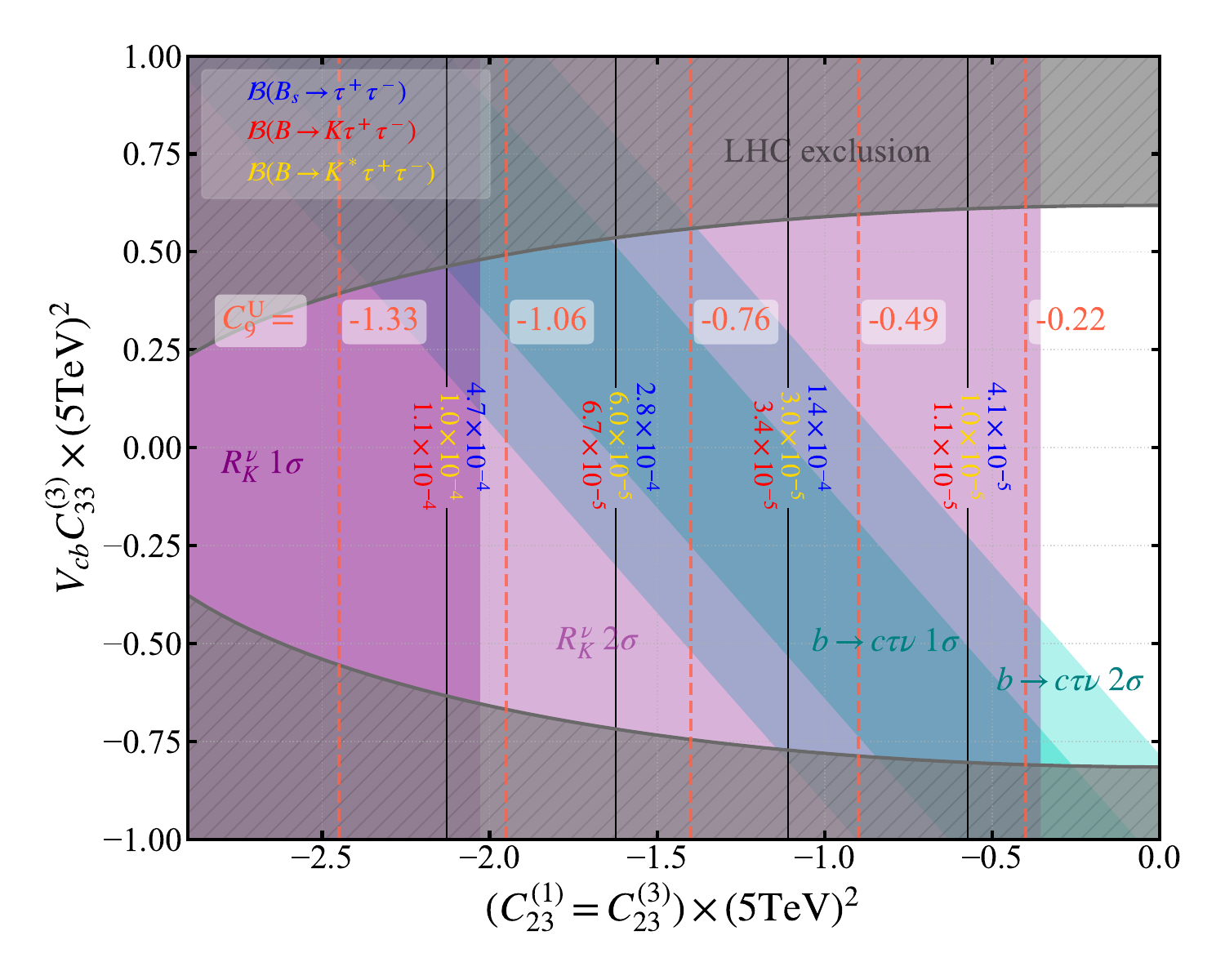}
    \end{subfigure}
    \hfill 
    \begin{subfigure}[b]
    {0.75\textwidth}
        \centering
        \includegraphics[width=\textwidth]{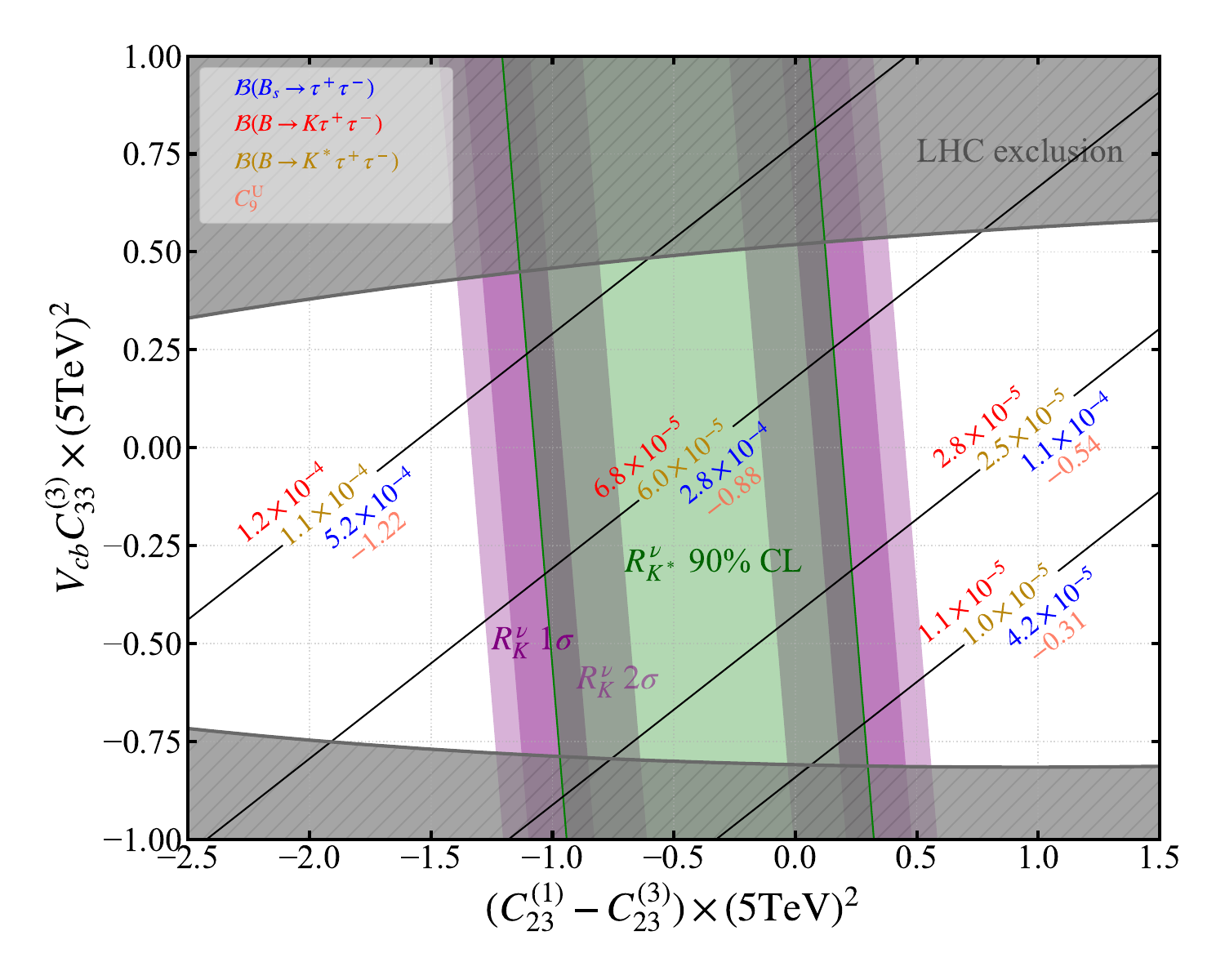}
    \end{subfigure}
    \caption{ Preferred and excluded regions in the $\left(C_{23}^{(1)}=C_{23}^{(3)}\right)$--$V_{cb}C_{33}^{(3)}$ plane (top) and in the $\left(C_{23}^{(1)}-C_{23}^{(3)}\right)$--$V_{cb}C_{33}^{(3)}$ plane (bottom) for a NP scale $\Lambda=5$\,TeV. The diagonal cyan band in the top plot is preferred by $b\to c\tau\nu$ measurements, while in the bottom panel, $R(D^{(*)})/R(D^{(*)})_{\text{SM}}=1.17$ is assumed. The purple regions are preferred by $R_K^\nu$, while the green region in the bottom panel is allowed by $R_{K^*}^\nu$ at $90\%$ CL (in the top panel, the whole plotted region is allowed by $R_{K^*}^\nu$). The hatched regions are excluded by non-resonant mono-tau searches at the LHC. The black lines show the central values for $\mathcal{B}(B_s\to \tau^+\tau^-)$, $\mathcal{B}(B\to K^{(*)}\tau^+\tau^-)$, and $C_9^\text{U}$.}
    \label{fig:scn4and5}
    \end{figure}
    
So far, we have considered the case of a generic flavour structure for the Wilson coefficients. However, if there is a hierarchical structure, $C_{33}^{(3)}\gg C_{23}^{(3)}$, there is the additional contribution via CKM rotations in the charged current (working in the down basis)
\begin{equation} \label{eq: lagrangian scenario 4}
\mathcal{L}_{\text{eff}}^{\text{NP}}\supset 
    2C_{33}^{(3)}V_{cb}(\bar{c}_L\gamma_\mu b_L)(\bar{\tau}_L\gamma^\mu\nu_{L\tau})\,.
\end{equation}
This means that the one-to-one correlation between the charged $b\to c\tau\nu$ current and the neutral $b\to s\tau^+\tau^-$ and $b\to s\nu\bar{\nu}$ currents is broken. Depending on the sign of $C_{33}^{(3)}$ w.r.t.~$C_{23}^{(3)}$, the relative effect in $R(D^{(*)})$ can be either enhanced or reduced, while all other relations we considered before stay unchanged (e.g.~the inclusion of right-handed currents and the relation of $b\to s\tau^+\tau^-$ processes with $C_9^\text{U}$). However, different balances between $C_{23}^{(1)}$ and $C_{23}^{(3)}$ to satisfy $R(D^{(*)})$ imply different effects on the neutral currents.

We show this scenario in Fig.~\ref{fig:scn4and5} for the case of $C_{23}^{(1)}=C_{23}^{(3)}$ (top) and $C_{23}^{(1)}\neq C_{23}^{(3)}$ (bottom). For the latter case, we fix $R(D^{(*)})/R(D^{(*)})_{\rm SM}=1.17$ to determine $C_{23}^{(3)}$. One can see that this scenario is more constrained by LHC searches such that very hierarchical structures are not able to fully explain $b\to c\tau\nu$ data. Regarding effects on the neutrino modes, one can see in the top panel that large positive allowed values of $C_{33}^{(3)}$ could lead to an alternative to the $C^{(1)}_{23}\neq C^{(3)}_{23}$ condition for enhancing $B\to K^{(*)}\nu\bar{\nu}$. However, the hierarchical structure alone is insufficient to account for the current central value of $R_K^\nu$. Moreover, in a potential future scenario where the $b\to s\tau^+\tau^-$ modes point towards an overcompensation of the SM in $R_K^\nu$, giving up the condition $C^{(1)}_{23}=C^{(3)}_{23}$ would become unavoidable and hierarchical structures as an alternative explanation alone would not be viable anymore. This can be seen in the bottom panel, where the scenario $C^{(1)}_{23}=C^{(3)}_{23}$ can be compatible with the solution for $R_K^\nu$ with NP constructively interfering with the SM (right purple band), but is incompatible with the solution with NP overcompensating the SM (left purple band).

\section{Conclusions}\label{sec:conclusions}

In this work, we have presented a comprehensive analysis of tauonic FCNC $B$-meson decays. We provided semi-analytic expressions for $B\to K^{(*)}\tau^+\tau^-$ and $B_s\to \phi\tau^+\tau^-$ over the full 
kinematically accessible $q^2$ range, including dominant $\psi(2S)$ and NP effects parametrised in the WET. This provides theory predictions directly comparable to measurements at LHCb and CMS, where $q^2$ reconstruction is difficult.

We then performed a detailed SMEFT analysis, exploring several flavour structures and including both left- and right-handed quark currents in light of the anomalies in $R(D^{(*)})$ and $B\to K^{(*)}\nu\bar\nu$. The correlations implied by $SU(2)_L$ gauge invariance, which link $b\to s\tau^+\tau^-$ transitions to the charged-current observables $R(D^{(*)})$ and the neutrino modes $B\to K^{(*)}\nu\bar\nu$, lead to large amplification factors, such that an $\mathcal{O}(10\%)$ effect in $R(D^{(*)})$ can result in an enhancement of several orders of magnitude in $b\to s\tau^+\tau^-$ branching ratios. We demonstrated that due to the constructive or destructive interference with the SM in $b\to s\nu\bar\nu$ processes, different NP scenarios can predict the same values for $\mathcal{B}(B\to K^{(*)}\nu\bar\nu)$. However, these scenarios can be disentangled by measuring $B_s\to(\phi)\tau^+\tau^-$ and $B\to K^{(*)}\tau^+\tau^-$.

Overall, our results highlight that $b\to s\tau^+\tau^-$ decays are a sensitive probe of NP connected to third-generation quarks and leptons. With LHCb and CMS entering high-luminosity operation, these modes provide promising search channels for which a significant signal at the expected sensitivities would constitute a clear evidence of NP.

\section*{Acknowledgements}

We thank Kostas Petridis,  Mitesh Patel and Hanae Tilquin
for very useful discussions. J.M.~gratefully acknowledges the financial support from ICREA under the ICREA Academia programme  
and from AGAUR under the ICREA Academia programme 2024. G.B.~, J.M.~ and M.N.~ acknowledge financial support from the Spanish Ministry of Science, Innovation and Universities (Agencia Estatal de Investigaci\'on MCIUI/AEI/10.13039/501100011033) through the Severo Ochoa Centers of Excellence Programme under Grants No. CEX2024-001441-S and No.~CEX2023-001292-S, and through Grants No. PID2023-146142NB-I00  and No.~PID2023-146220NB-I00. J.M.~is a Serra Húnter Fellow.

\bibliographystyle{JHEP}
\bibliography{main.bib}
\end{document}